\documentclass[aps,a4paper,twocolumn]{revtex4}

\usepackage{amsmath}
\usepackage{graphicx}
\usepackage{subfigure}
\usepackage[usenames,dvipsnames]{color}
\definecolor{darkblue}{RGB}{0,0,196}
\usepackage[colorlinks=true,linkcolor=darkblue,citecolor=darkblue,urlcolor=darkblue]{hyperref}
\usepackage{setspace}
\usepackage{multirow}
\usepackage{hyperref}
\usepackage{xcolor}
\usepackage{xspace}
\hypersetup{
  colorlinks   = true, 
  urlcolor     = blue, 
  linkcolor    = blue, 
  citecolor   = blue 
}
\usepackage{graphicx}
\usepackage{amsmath,bbm}
\usepackage{amssymb,bm}
\usepackage{yfonts}
\usepackage{comment}
\usepackage[normalem]{ulem}

\newcommand{\dvdy}{\ensuremath{{\rm d}V/{\rm d}y}\xspace}

\begin{document}

\title{Confronting the production mechanisms of nuclei with deuteron and proton-triggered balance functions}
\author{Sushanta Tripathy\footnote{sushanta.tripathy@cern.ch}}
\author{Peter Christiansen\footnote{peter.christiansen@cern.ch}}

\affiliation{Division of Particle and Nuclear Physics, Department of Physics, Lund University, Lund, 22364, Sweden}

\date{\today}

\begin{abstract}
In ultra high-energy collisions, nuclei with very low binding energies are not expected to survive the dense and hot final state environment. The traditional view has therefore been that nuclei form via coalescence after the hot environment has dissipated. However, statistical thermal models, where hadrons are produced from a fireball at thermal equilibrium, can describe the relative abundances of light nuclei in pp and heavy-ion collisions at the LHC equally well. \\ 
In this paper we investigate if balance functions triggered by protons and deuterons can be used to distinguish between the two production mechanisms. The coalescence model is investigated using PYTHIA, while the statistical thermal model is examined using the Thermal FIST package. We find that for both models the same simple relation between proton and deuteron triggered balance functions is applicable. However, there is a striking difference between the two models when the transverse momentum of trigger particles is varied. This dependence offers a promising observable to discriminate between the two production scenarios that goes beyond nuclei production. Furthermore, we find that deuteron–meson balance functions vanish identically for both models due to baryon number conservation and isospin symmetry.
\end{abstract}
  
\maketitle


\section{Introduction}
\label{intro}
The Large Hadron Collider (LHC) at CERN and the Relativistic Heavy-Ion Collider (RHIC) at BNL have rich physics programs focused on quantum chromodynamics (QCD). Due to its strong coupling limit, a fundamental QCD understanding is often lacking even for simple basic observables, for example how quarks and gluons form hadrons (hadronization). Experimental measurements therefore often drive new breakthroughs. A hot topic at the LHC is the production of nuclei (comprising of two or more baryons) and hypernuclei (comprising of a proton, a neutron, and a lambda hyperon), which has been extensively studied in the last two decades~\cite{nuclei_pp_PbPb,nuclei_pp,deuteron_pp_7TeV,deuteron_pPbALICE,3He_pPb,deuteron_pp_13TeV,nuclei_pp_5TeV,nuclei_pp_13TeV_HM}. In ultra high-energy lead-lead (Pb--Pb) and proton-proton (pp) collisions at the LHC, composite objects like nuclei with very low binding energies (for instance, the deuteron binding energy is ${\approx}$2.2\,MeV), are not expected to survive the dense and hot final-state environment. As this is an open problem, the popular event generators in the field, such as PYTHIA~\cite{Bierlich:2022pfr} and EPOS~\cite{Pierog:2013ria}, do not include the production of nuclei in their default configuration. This production mechanism of nuclei is also relevant in astrophysics where antinuclei are used to probe dark matter by space-borne experiments~\cite{Korsmeier:2017xzj,Blum:2017qnn,Kachelriess:2020uoh,vonDoetinchem:2020vbj,PhysRevD.105.083021}, and the antinuclei produced by hadronic collisions in space constitutes an irreducible background. 

The traditional view has been that nuclei and antinuclei form via coalescence~\cite{Sato:1981ez,Nagle:1996vp,Scheibl:1998tk,Blum:2017qnn,Blum:2019suo,Mrowczynski:2019yrr,Bellini:2020cbj,Mahlein:2023fmx}, after the hot environment has dissipated. However, statistical thermal models, where hadrons are produced from a fireball at thermal equilibrium, can describe the relative abundances of light nuclei in pp and heavy-ion collisions at the LHC~\cite{SHM1,SHM2,SHM3,SHM4,SHM5,SHM6}. At the LHC, where antimatter-over-matter ratios are found to be unity, these statistical thermal models have only one free parameter, the hadronization temperature T, to describe all relative light-flavour hadron (with $u$, $d$, $s$ valence quarks) abundances and even get the antihyperons spot on~\cite{SHM5}. Nuclei and antinuclei produced via statistical thermal models have coined the term ``Snowballs in Hell” because the temperature used to describe the yields is ${\approx}$160\,MeV while the binding energies are of order a few MeV. The prediction of the relative yields by the statistical thermal models could be accidental, but could also be fundamental and provide completely new and unexpected insights into QCD and the hadronization process in particular.
Thus, the big challenge is to come up with fundamentally new measurements that can probe the microscopic production of nuclei and thereby resolve this ``nuclei puzzle''. 

Balance functions were originally proposed as sensitive probes of hadronization time in heavy-ion collisions~\cite{Bass:2000az}. Recently, there has also been a big interest in pp collisions with ALICE measurements of $\Xi$-triggered  balance functions in pp collisions~\cite{ALICE:2023asw} and phenomenological studies~\cite{Pruneau:2024jpa,Bierlich:2025pkg}. These new papers have demonstrated that balance functions can be highly sensitive to the hadronization mechanism in pp collisions. In a similar spirit, we investigate deuteron- and proton-triggered balance functions in rapidity and azimuthal angle. 

Balance functions probe the quantum number balance and in this way get additional insights into the underlying hadronization mechanism. By measuring the relative distributions of ``associated'' hadrons (protons, $\Lambda$ and pions) with opposite baryon number ($B$) with respect to a ``trigger'' deuteron and proton, the distribution of the balancing quantum numbers in momentum space can be determined. In pp collisions at LHC, $B \approx 0$ for measurements done at mid-rapidity and so a deuteron ($B = 2$) will be balanced by $B = -2$ baryons. If the deuteron is formed via the coalescence of a proton and a neutron that are close in momentum space, then the balance of the deuteron will be the product of the balance of the proton and neutron. If on the other hand, the deuteron is produced directly in the collision, then there is no reason for this to be the case as the baryon number only has to be conserved globally by the thermal system. We therefore want to understand, in particular, the relation between the deuteron and proton balance functions to test their sensitivity to these differences. In addition, due to isospin symmetry, the balance of deuterons with neutrons is expected to be identical to that with protons. Thus, a deuteron with electric charge $Q = 1$ could, in principle, be balanced by two neutrons ($Q = 0$), but such a process would necessarily require a negatively charged meson to conserve charge. Since neutron balance cannot be measured experimentally, we instead study the balance with pions, alongside protons and $\Lambda$ baryons. Furthermore, the balance will also be studied as a function of transverse momentum and multiplicity, to test how this affects the discrimination power. This article presents a proof-of-concept and focuses on the studies in pp collisions at $\sqrt{s} = 13$\,TeV, which can be extended to collisions with larger system size and/or different collision energy.

\begin{figure}
    \centering
\includegraphics[width=1.05\linewidth]{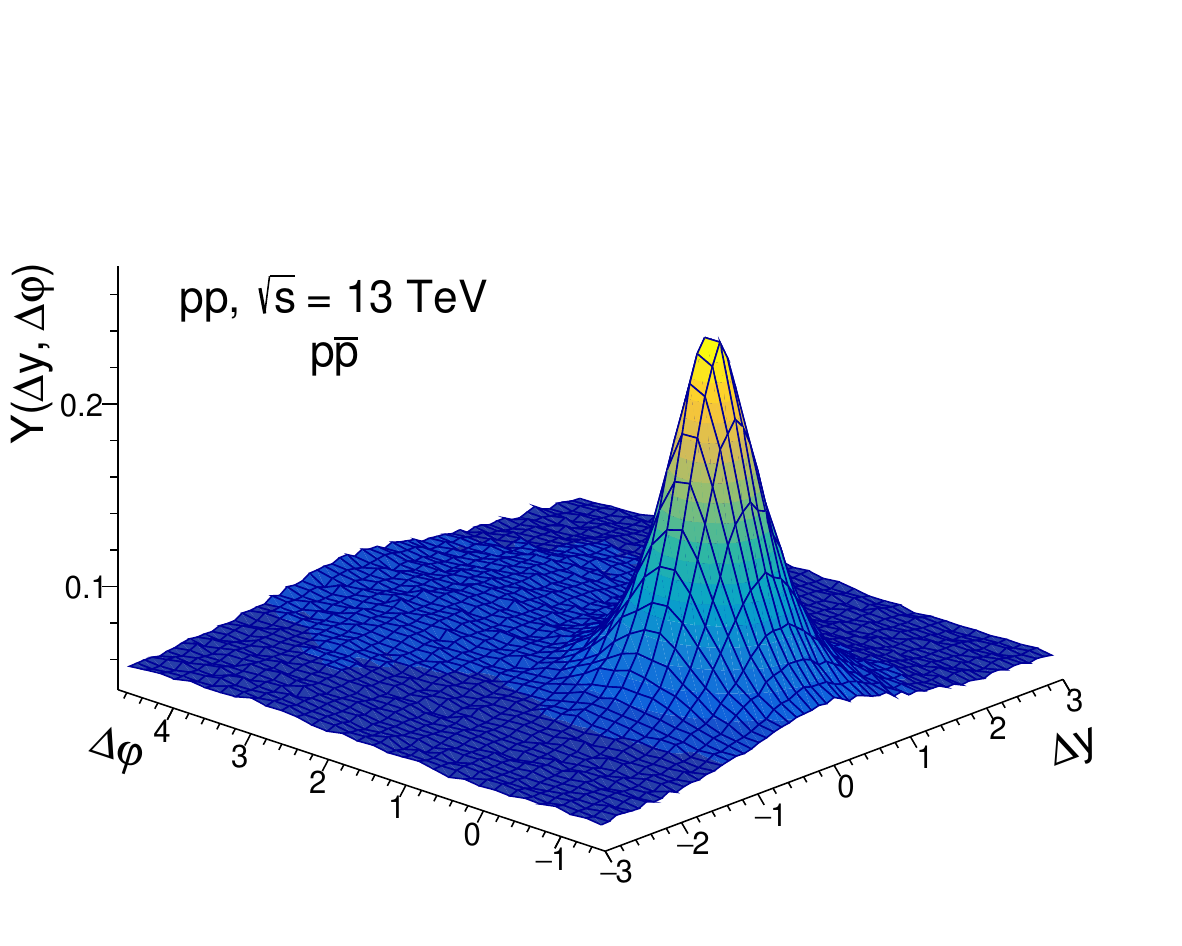}
    \caption{ A representative plot of the associated yield of antiprotons per triggered protons in pp collisions at $\sqrt{s}$ = 13 TeV.}
\label{fig:2dcorrelation}
\end{figure}

The paper is organized as follows. In Sec.~\ref{methods_model}, the deuteron and proton balance functions are introduced and it is explained how they are obtained from the PYTHIA and Thermal FIST models. The results are presented and discussed in Sec.~\ref{results}, followed by the conclusions in Sec.~\ref{conclusion}.

\section{Methodology and Models}
\label{methods_model}

\begin{figure}
    \centering
\includegraphics[width=1\linewidth]{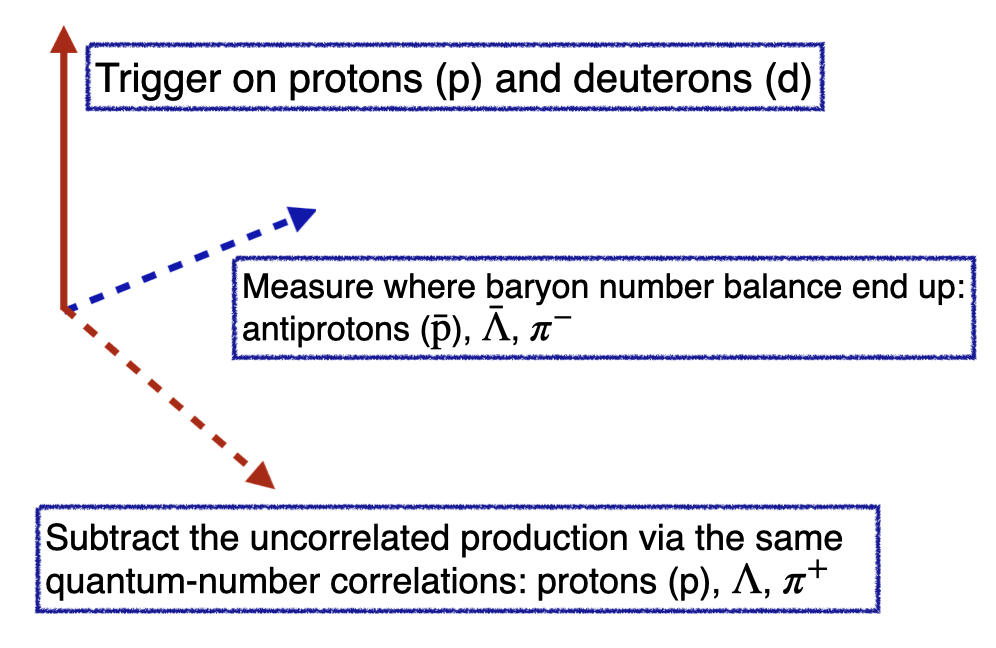}
    \caption{Depiction of analysis methodology to estimate the proton and deuteron-triggered balance functions}
\label{fig:AnalysisMethod}
\end{figure}

As discussed briefly in the previous section, we propose a new set of measurements of how nuclei are balanced by protons, Lambda ($\Lambda$) baryons and pions in pp collisions at the LHC. For this purpose, the associated yield per trigger particle needs to be obtained which is defined as,
\begin{equation}
Y(\Delta y, \Delta \varphi) = \frac{1}{N_{\text{trig}}} \frac{d^2 N_{\text{pairs}}}{d\Delta y\, d\Delta \varphi},
\end{equation}
where $\Delta \varphi = \varphi_{\text{assoc}} - \varphi_{\text{trig}}$ is the relative azimuthal angle and $\Delta y = y_{\text{assoc}} - y_{\text{trig}}$ is the difference in rapidity between the trigger and associated particles. $N_{\text{trig}}$ is the number of trigger particles and $N_{\text{pairs}}$ is the number of trigger–associated particle pairs. For the trigger particles, we restrict the pseudorapidity range,  $|\eta| < 0.8$. In the current study, we avoid the usage of an acceptance cut on the associated particles. In this way, there is no need to perform event-mixing corrections for the restricted acceptance and we recover the full quantum-number balance. However, in experiments, the detector acceptance and response need to be taken into account and thus, event-mixing corrections need to be performed. Although the balance function can be studied as a function of both rapidity ($\Delta y$) and azimuthal angle ($\Delta\varphi$), here we focus on the rapidity dependence as one of the models (Thermal FIST) studied in this paper does not generate realistic azimuthal correlations. Throughout this paper, the balance function is denoted as $B(\Delta y)$. Figure~\ref{fig:2dcorrelation} shows an example of $Y(\Delta y, \Delta \varphi)$ where the trigger is a proton or antiproton and the associated particle is an opposite-baryon-number proton or antiproton, respectively. In order to isolate the quantum-number-balancing part of  $Y(\Delta y, \Delta \varphi)$ and remove correlations due to (mini)jet fragmentation or the underlying event, the difference between the opposite quantum-number (``opposite-sign'') and same-quantum-number (``same-sign'') correlations is also calculated, see Fig.~\ref{fig:AnalysisMethod} for an illustration. The balance function for deuterons (trigger) with anti-protons is therefore defined as:
\begin{equation}
B_{d\bar{p}}(\Delta y) = Y_{d\bar{p}}(\Delta y) - Y_{dp}(\Delta y).
\end{equation}
Finally note that the balance functions shown in the following are averaged over particles and antiparticles as they are produced in equal numbers at the LHC energies we discuss here. For more details on the balance functions used here we refer to Refs.~\cite{ALICE:2023asw,Bierlich:2025pkg}.

The PYTHIA event generator and Thermal FIST model have fundamentally different approaches to hadronization and particle production (including nuclei). In this way, they represent two contrasting paradigms: perturbative QCD-inspired string fragmentation versus thermal statistical hadronization, respectively. PYTHIA is a Monte Carlo event generator~\cite{Bierlich:2022pfr} inspired by perturbative QCD for partonic collisions, combined with the Lund string model for hadronization. In this framework, color fields between partons are represented as strings~\cite{Andersson:1983ia}, which fragment into hadrons through successive string breakings. In the default configuration of PYTHIA, deuteron coalescence is disabled; however, it can be enabled ($\texttt{HadronLevel:DeuteronProduction = on}$) to include deuteron production via coalescence. The coalescence implemented in PYTHIA is done in momentum space only. After the final state particles of an event are produced, protons and neutrons are selected and combined into pairs which may form deuterons. This is done via an empirical model~\cite{Dal:2015sha}, where cross-sections are used to determine if a combination binds into a deuteron. To ensure conservation of momentum and energy, the final state for each deuteron production channel is required to have at least two products, where one product is a deuteron. For this study, we have used PYTHIA version 8.3 with the default Monash tune.

Thermal FIST (Fast Implementation of Statistical Thermodynamics)~\cite{Vovchenko:2019pjl} is a statistical hadronization model designed to describe particle production under the assumption of thermal equilibrium. It implements a hadron resonance gas framework, where particle yields and fluctuations are governed by very few thermodynamic parameters: temperature, chemical potentials, and correlation volumes, and QCD only enters via the spectrum of hadrons that can be produced. In the context of this work, Thermal FIST is used to simulate event-by-event fluctuations based on global conservation laws and statistical distributions, without modeling the full, dynamical evolution of collisions. Importantly, Thermal FIST lacks the momentum-space correlations observed in pp collisions, particularly in azimuthal angle and $p_{\rm T}$, as it does not simulate particle production mechanisms but rather samples
from equilibrium distributions. Therefore the particles produced by the statistical thermal model are boosted using a Blast-Wave model that has been tuned to match the ALICE results for pp collisions at 13 TeV~\cite{ALICE:2020nkc}, which ensures that the shape of the $p_{\rm T}$ spectra are comparable to those measured by ALICE. A similar procedure was used in Refs.~\cite{ALICE:2024rnr,Bierlich:2025pkg}. The correlation length in Thermal FIST is controlled via the choice of correlation volume, $V_{\rm c}$, with smaller volumes leading to stronger local conservation effects. In the next section, we will use as default for $V_{\rm c}$ and $T$ the parameters needed to describe the ALICE strangeness enhancement~\cite{Vovchenko:2019kes}. In addition, two variations of model parameters will be shown: (1) the default $V_{\rm c} = 3\,\dvdy$, meaning that the correlation volume is three units of rapidity wide, and three other widths: $1\,\dvdy, 1.6\,\dvdy$, and $4.8\,\dvdy$. (2) the default $T$ value of 176\,MeV and a variation to 156\,MeV.

\section{Results and discussions}
\label{results}

\begin{figure*}[ht!]
\begin{center}
\includegraphics[scale=0.4]{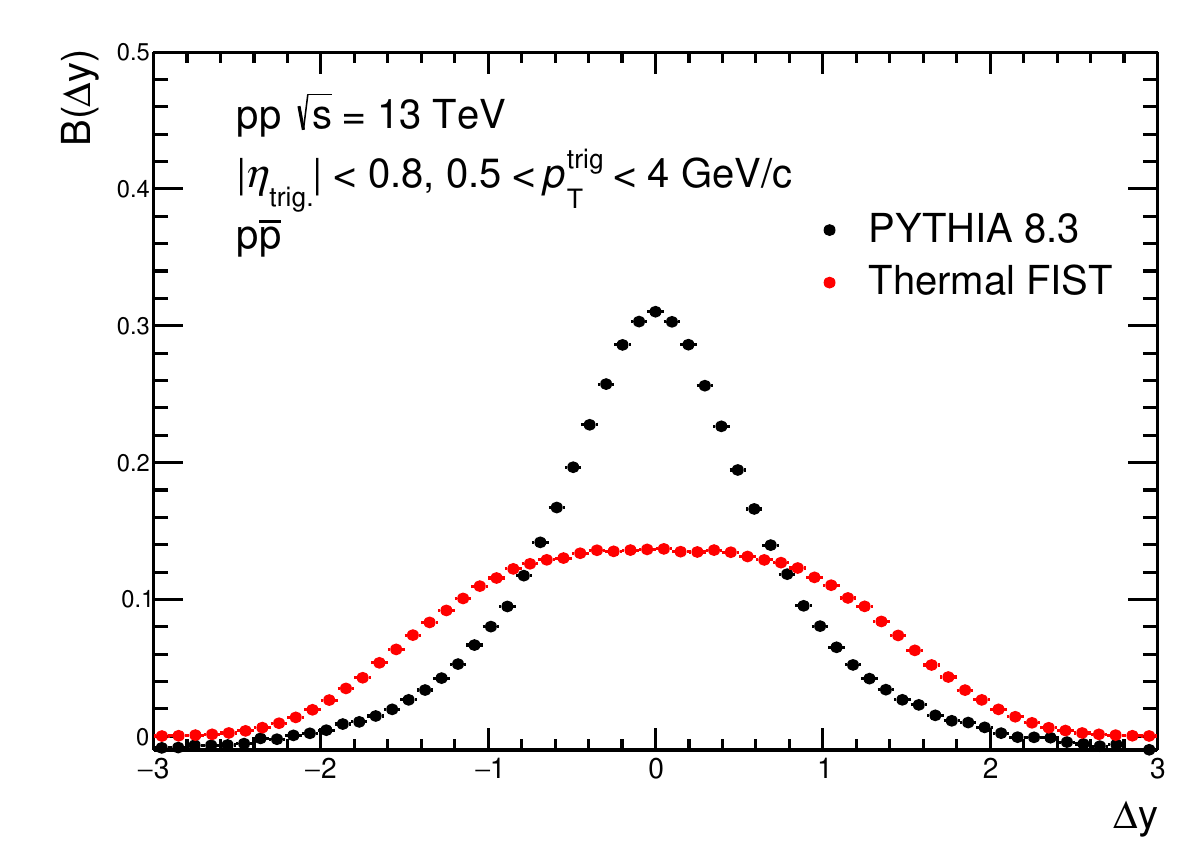}
\includegraphics[scale=0.4]{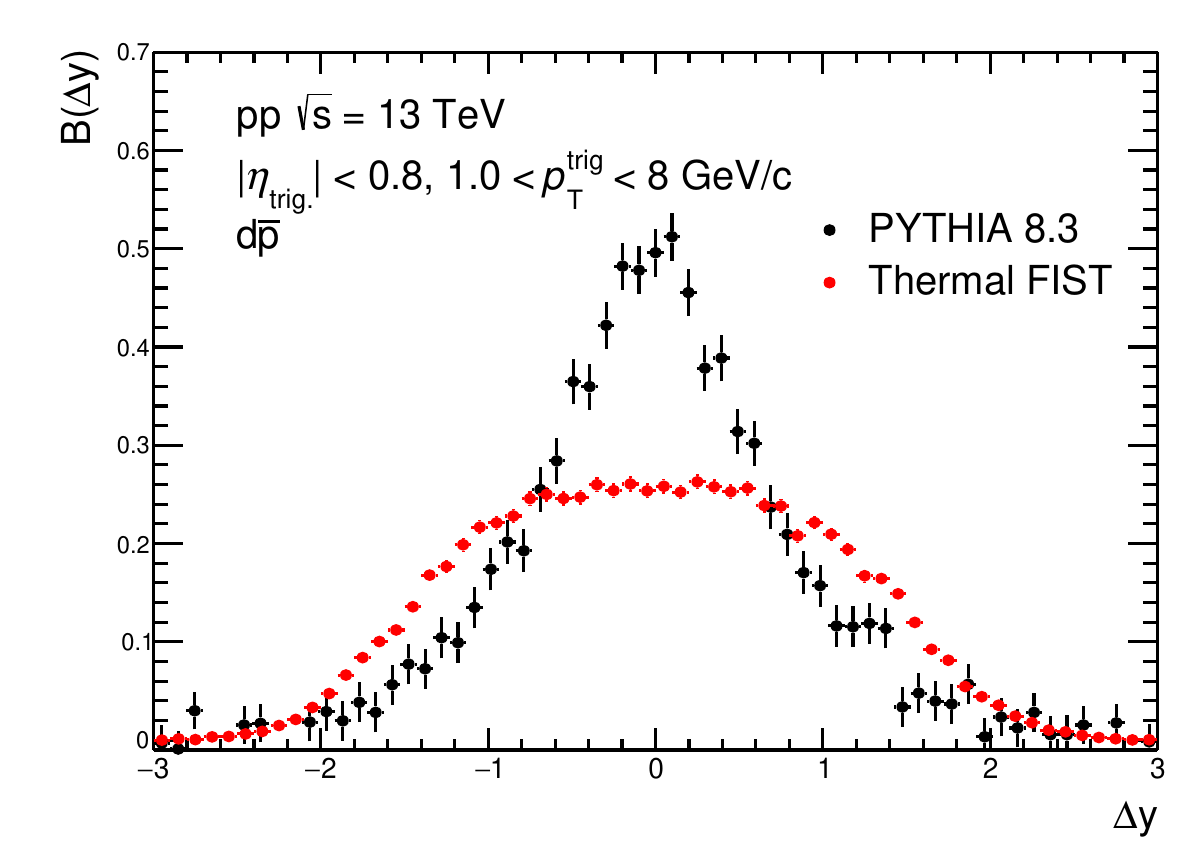}
\includegraphics[scale=0.4]{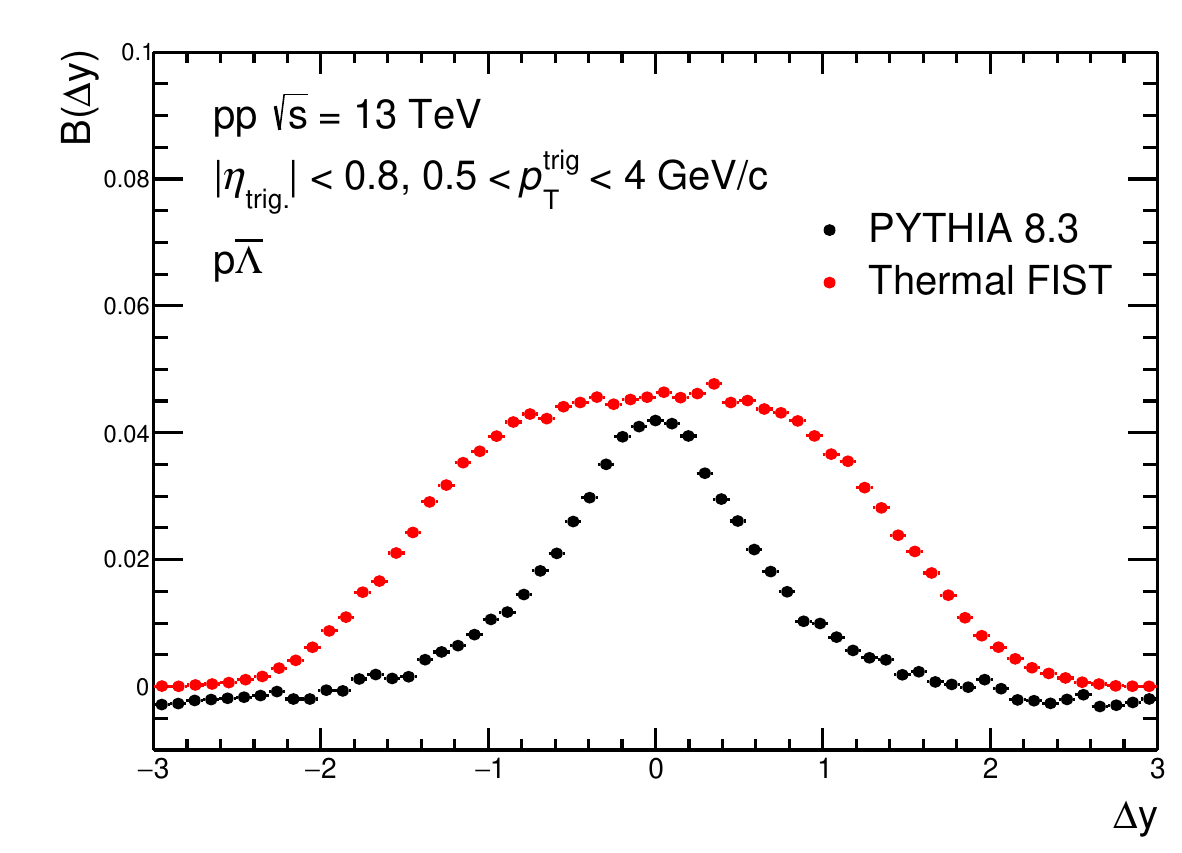}
\includegraphics[scale=0.4]{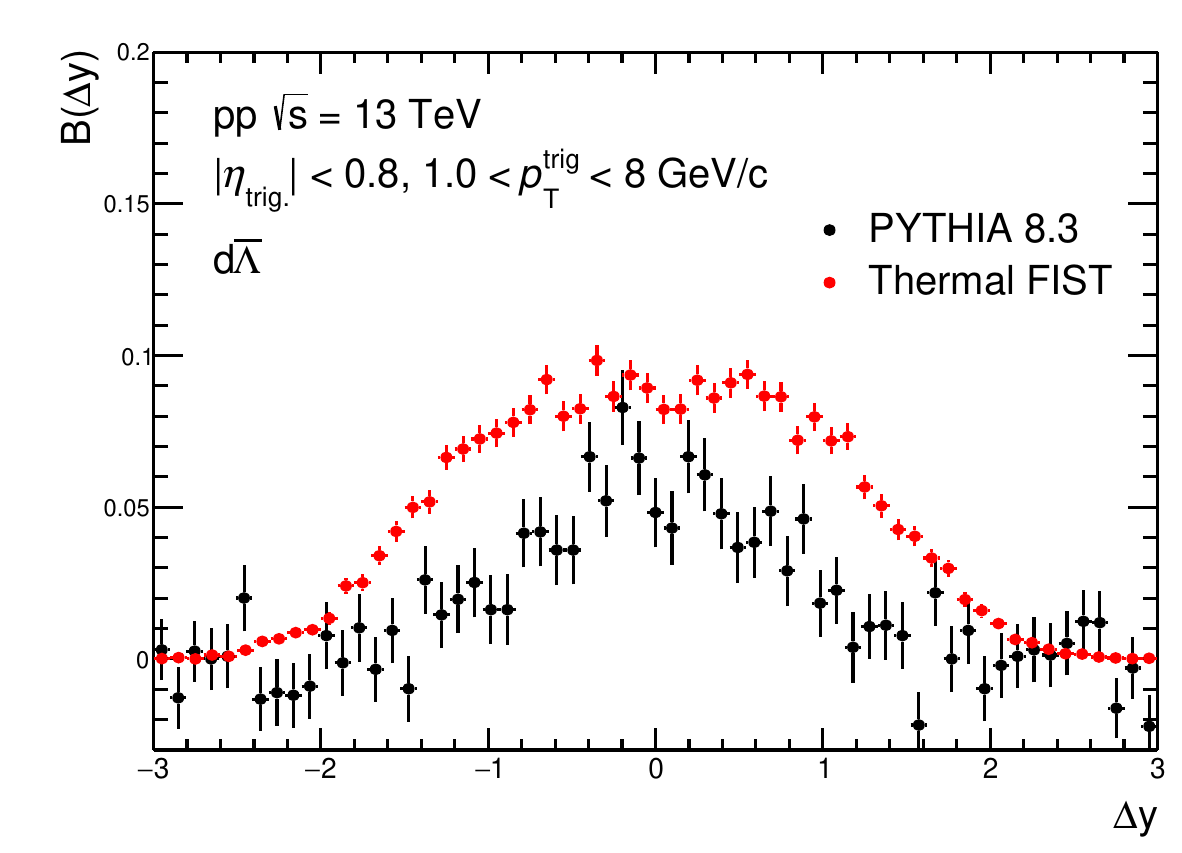}
\includegraphics[scale=0.4]{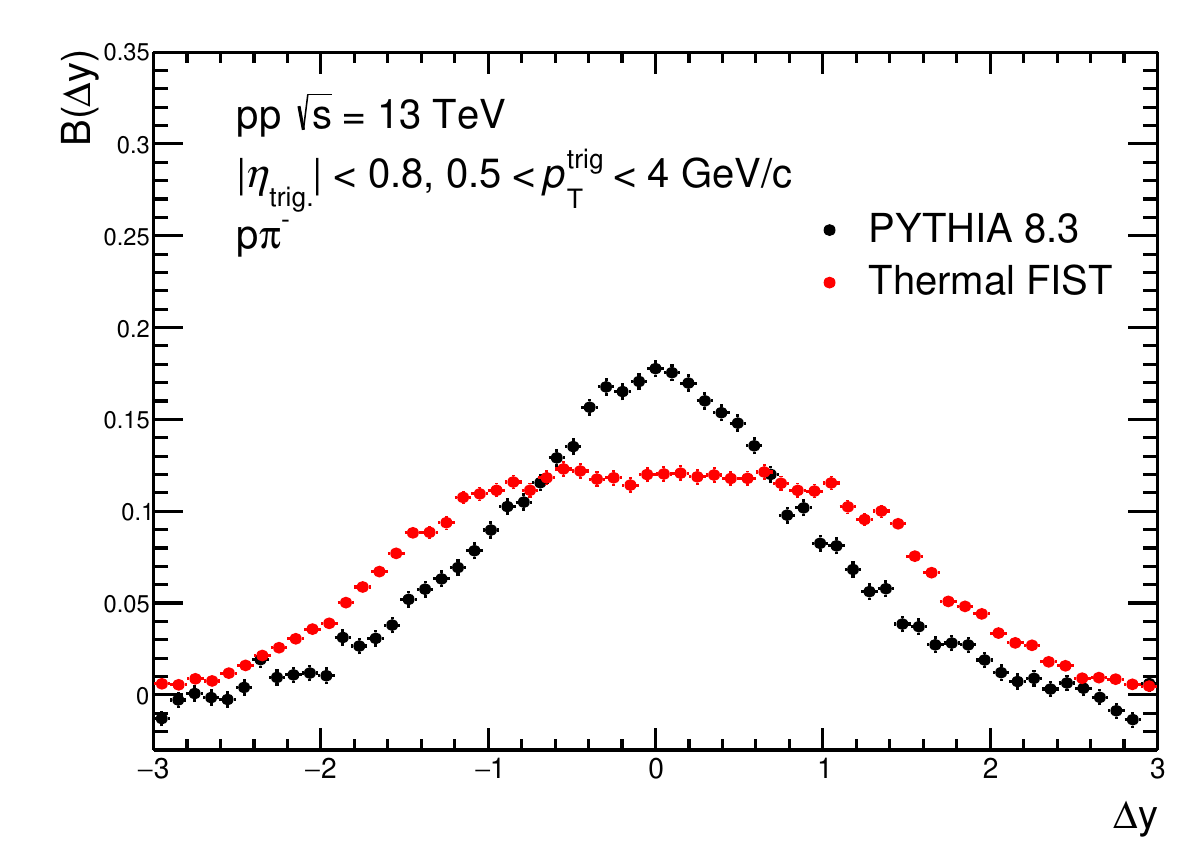}
\includegraphics[scale=0.4]{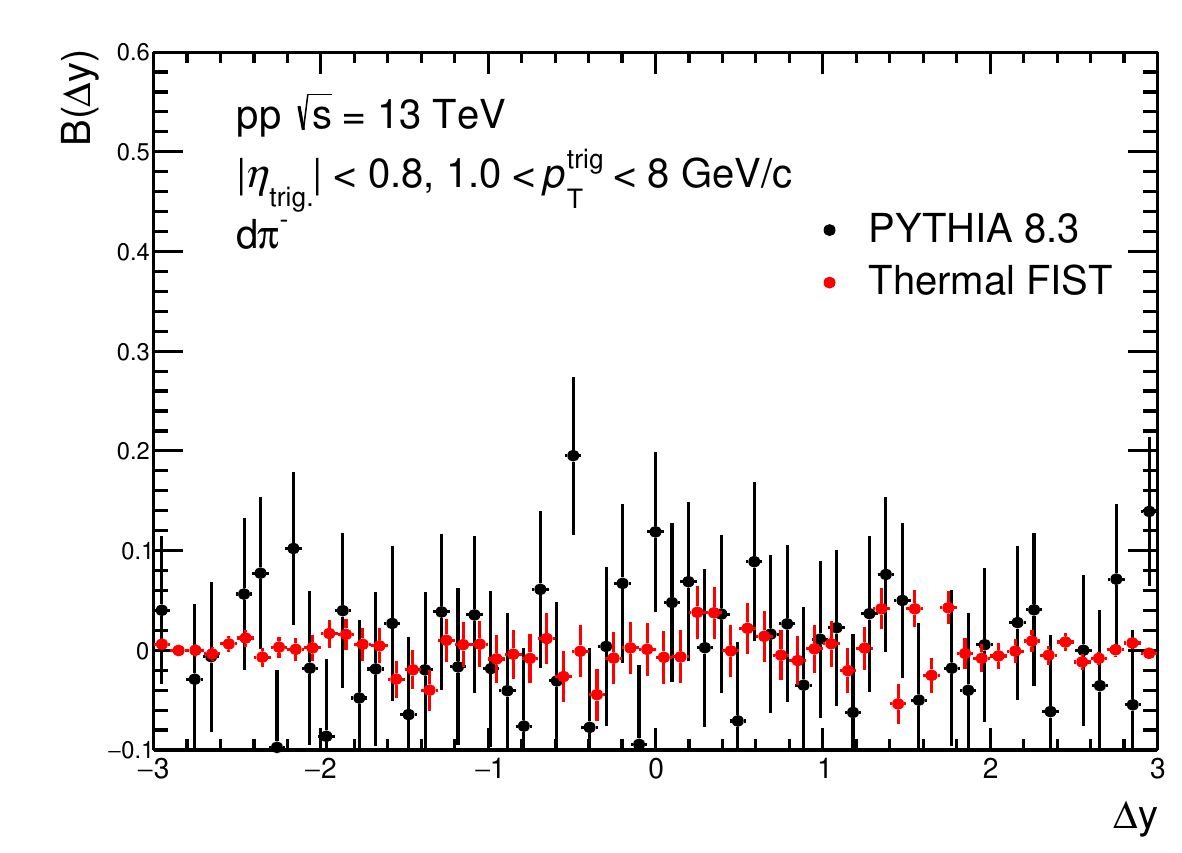}
\caption{Comparison of the balance function of triggered protons (left) and triggered deuterons (right) obtained from PYTHIA and Thermal FIST models.}
\label{fig:pp_dp_0p4GeV_model}
\end{center}
\end{figure*}
Having discussed the proposed analysis methodology and introduced the two models, the results will be presented in this section. 

The left and right panels of Fig.~\ref{fig:pp_dp_0p4GeV_model} show the balance functions for triggered protons and deuterons, respectively, obtained from PYTHIA and Thermal FIST. Triggered protons are selected within a transverse momentum range of $0.5$–$4$ GeV/$c$, while the deuterons are chosen in the range twice that of the protons, i.e., $1$–$8$ GeV/$c$. This choice follows the coalescence picture, in which nucleons must have similar and parallel momentum vectors to form a deuteron. In addition, this choice enables us to provide a prediction for possible future experimental measurements from ALICE as the chosen momentum ranges for trigger and associated particles are compatible with the particle identification capabilities of ALICE (using its Time Projection Chamber and Time-of-flight detector), see for example Ref.~\cite{ALICE:2020jsh}. There is no selection on the kinematic properties of associated particles, so the full balance of triggered protons and deuterons is recovered. 

At first glance, PYTHIA and Thermal FIST show visibly different balance function shapes for both protons and deuterons. In particular, Thermal FIST produces a broader distribution in $\Delta y$ compared to PYTHIA for both the $p\bar{p}$ and $d\bar{p}$ balance. This difference in broadness was also observed in Ref.~\cite{Bierlich:2025pkg} and will be examined in more detail later in the manuscript. In PYTHIA, mid-rapidity protons are mainly produced through string breakings, where diquark breakings produce diquark pairs $q_1q_2\bar{q_1}\bar{q_2}$. Deuterons are subsequently formed via coalescence, leading to balance functions for protons and deuterons with broadly similar features. A comparable shape is also observed for the balance by $\Lambda$ baryons, hinting at common production mechanisms. 

While looking at the balance with pions, we make an interesting observation. For the $p\pi^{-}$ balance, the behavior is similar to that of the $p\bar{p}$ and $p\bar{\Lambda}$ balances in both models. However, both PYTHIA and Thermal FIST predict a vanishing balance for $d\pi^{-}$. This can be understood from baryon number and charge conservation together with isospin symmetry. The process in which a proton is balanced by a $\pi^{-}$ and an antineutron have the same cross section as the one where a neutron is balanced by a $\pi^{+}$ and an antiproton. Since the deuteron consists of both a proton and a neutron, the two contributions cancel exactly, resulting in zero $d\pi^{-}$ balance. 

\begin{figure*}[ht!]
\begin{center}
\includegraphics[scale=0.4]{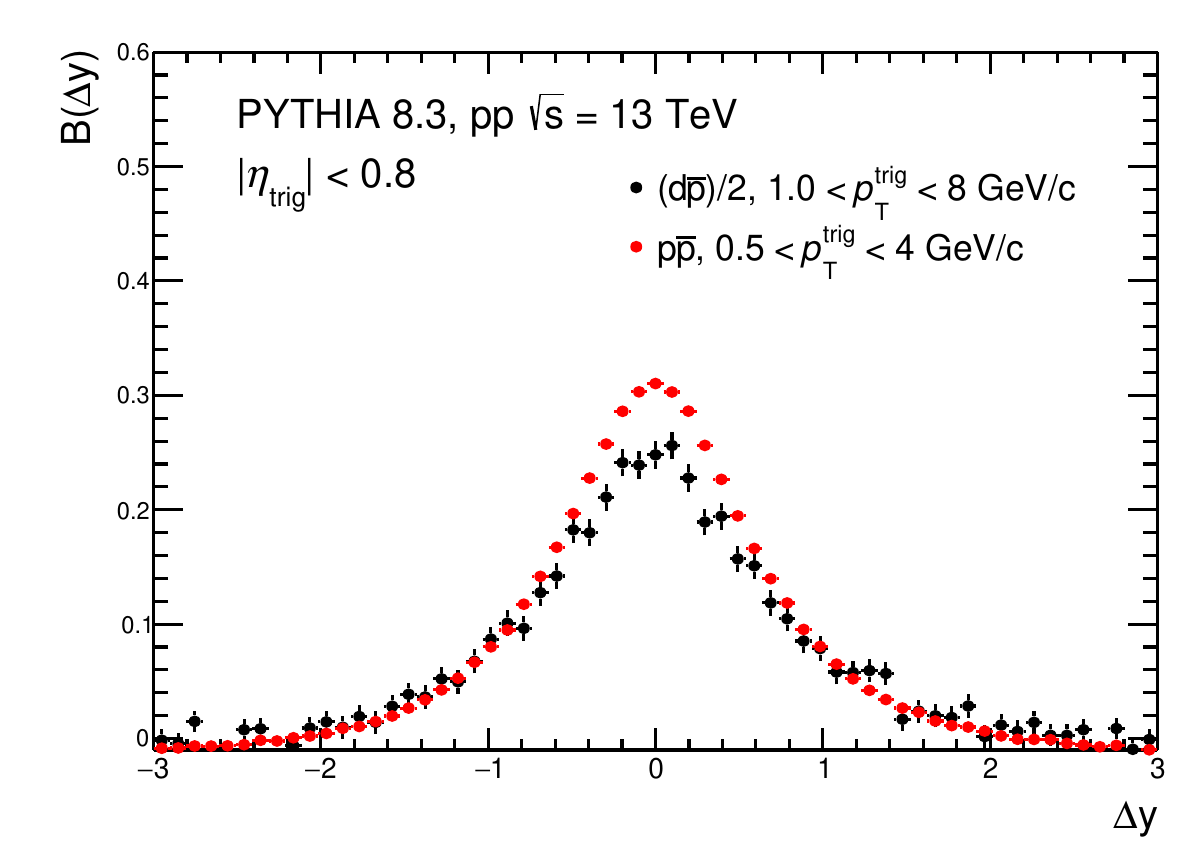}
\includegraphics[scale=0.4]{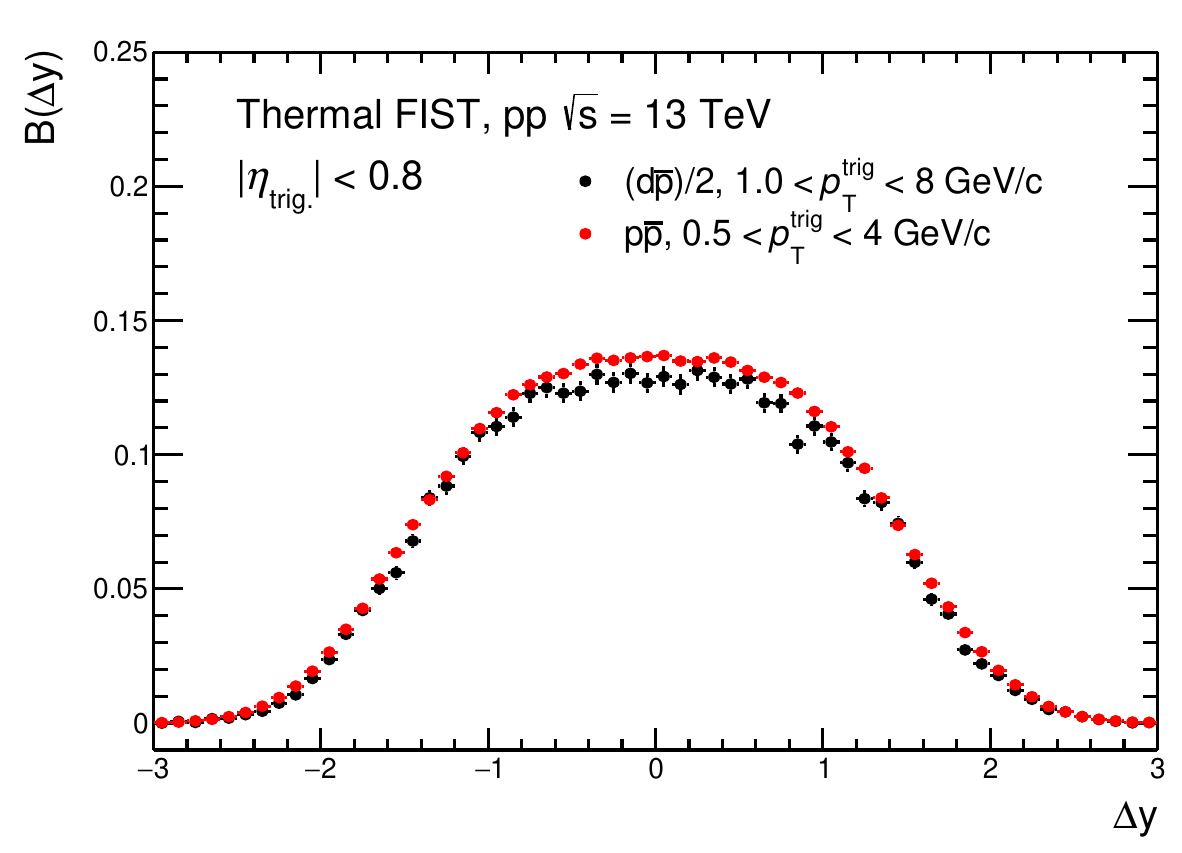}
\caption{Comparison of the balance function of triggered protons and triggered deuterons divided by two obtained from PYTHIA (left) and Thermal FIST (right) models.}
\label{fig:pp_dp_0p4GeV}
\end{center}
\end{figure*}

Figure~\ref{fig:pp_dp_0p4GeV} further compares the balance function of triggered protons with that of triggered deuterons scaled by a factor of $1/2$. Since a deuteron consists of two nucleons, its balance function is expected to be approximately twice that of a proton. Both models confirm this expectation, with the scaled deuteron balance functions largely overlapping the proton results. We have tested that the observed small difference agrees with that neutrons are slightly less likely to be balanced by antiprotons than protons.

\begin{figure*}[ht!]
\begin{center}
\includegraphics[scale=0.4]{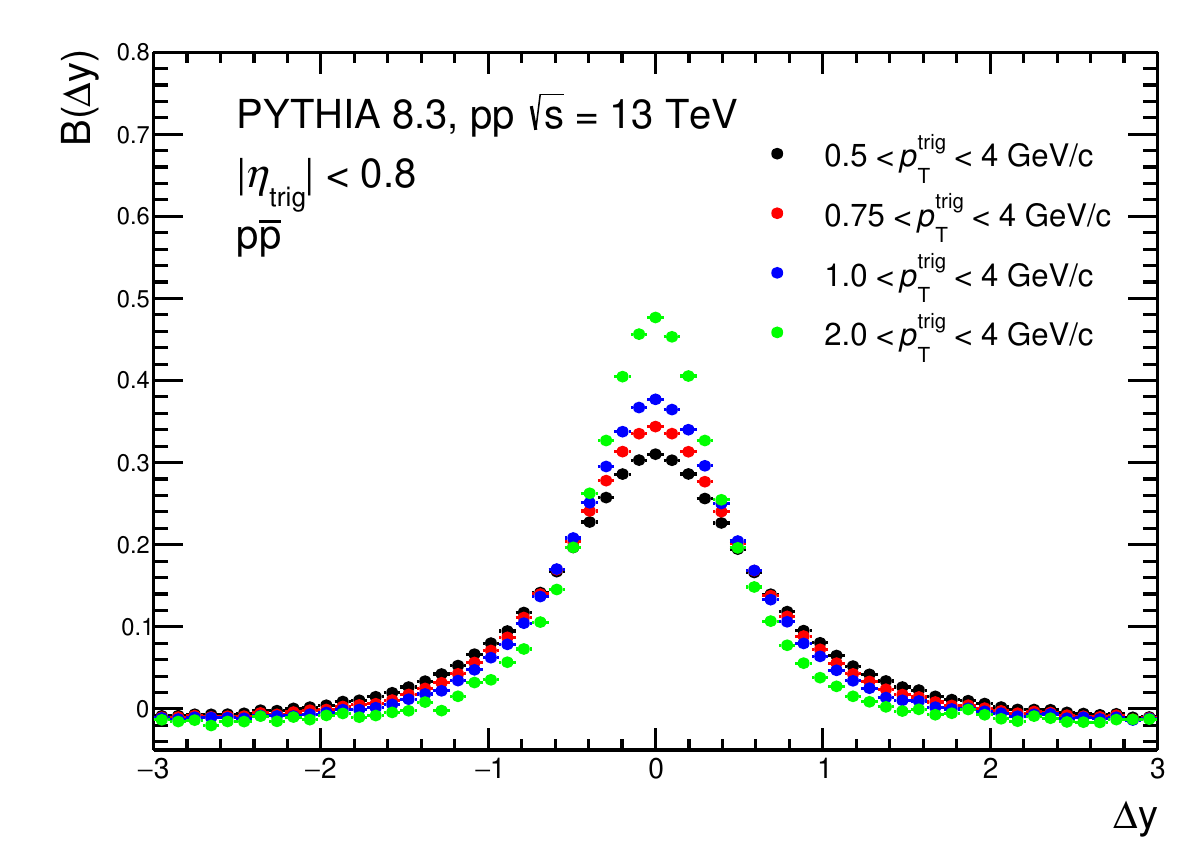}
\includegraphics[scale=0.4]{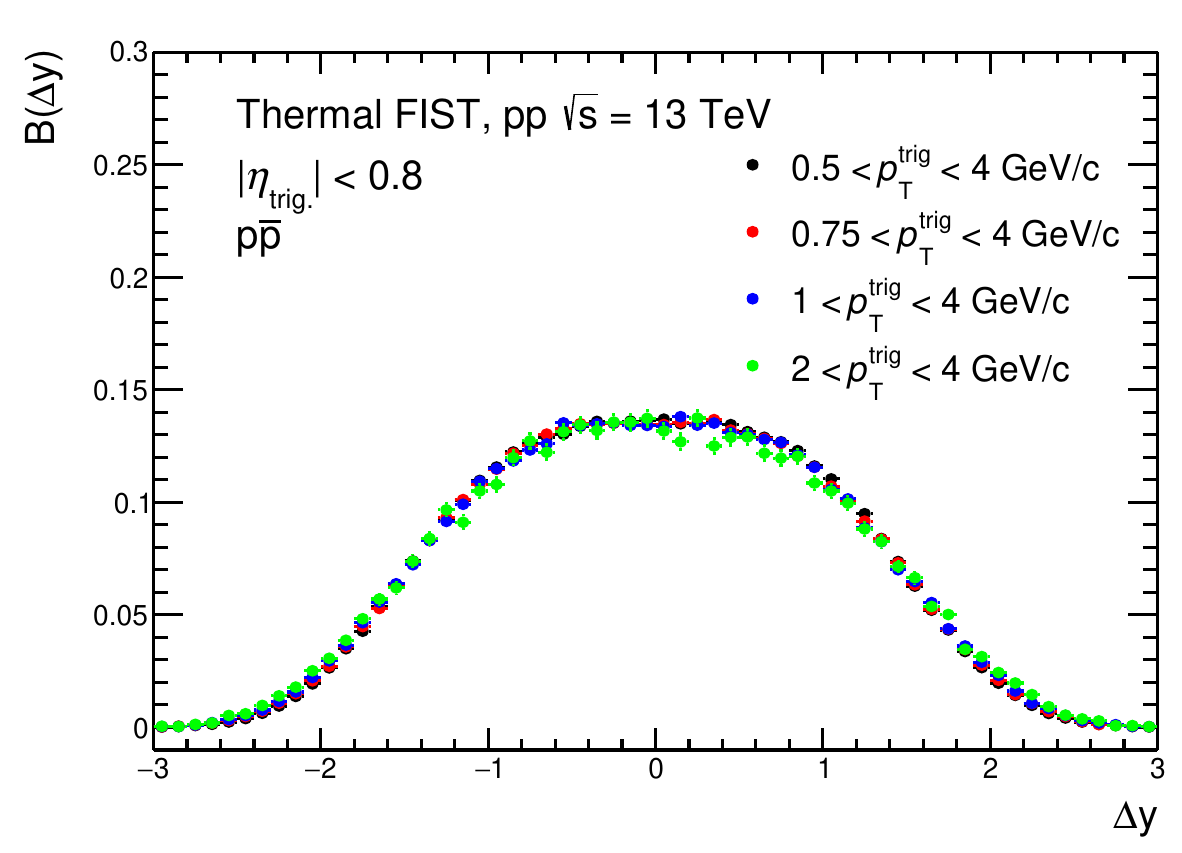}
\includegraphics[scale=0.4]{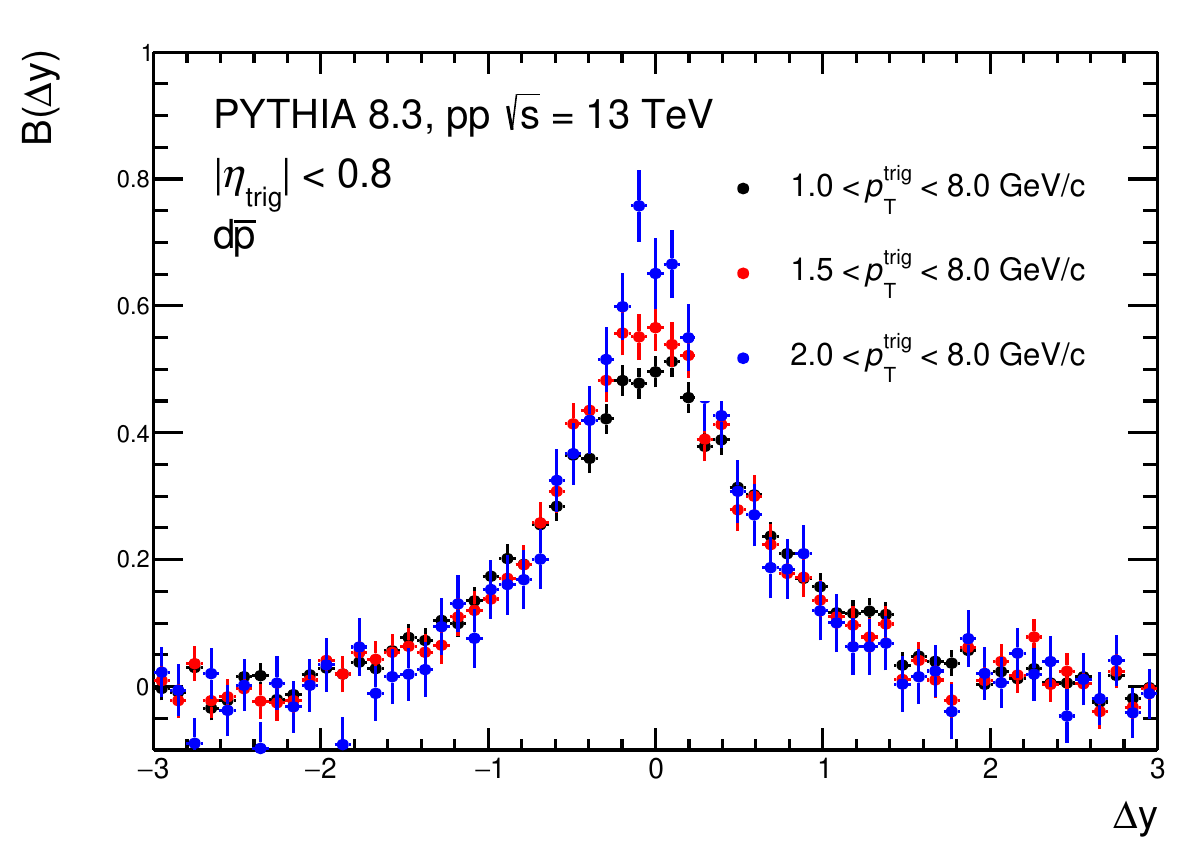}
\includegraphics[scale=0.4]{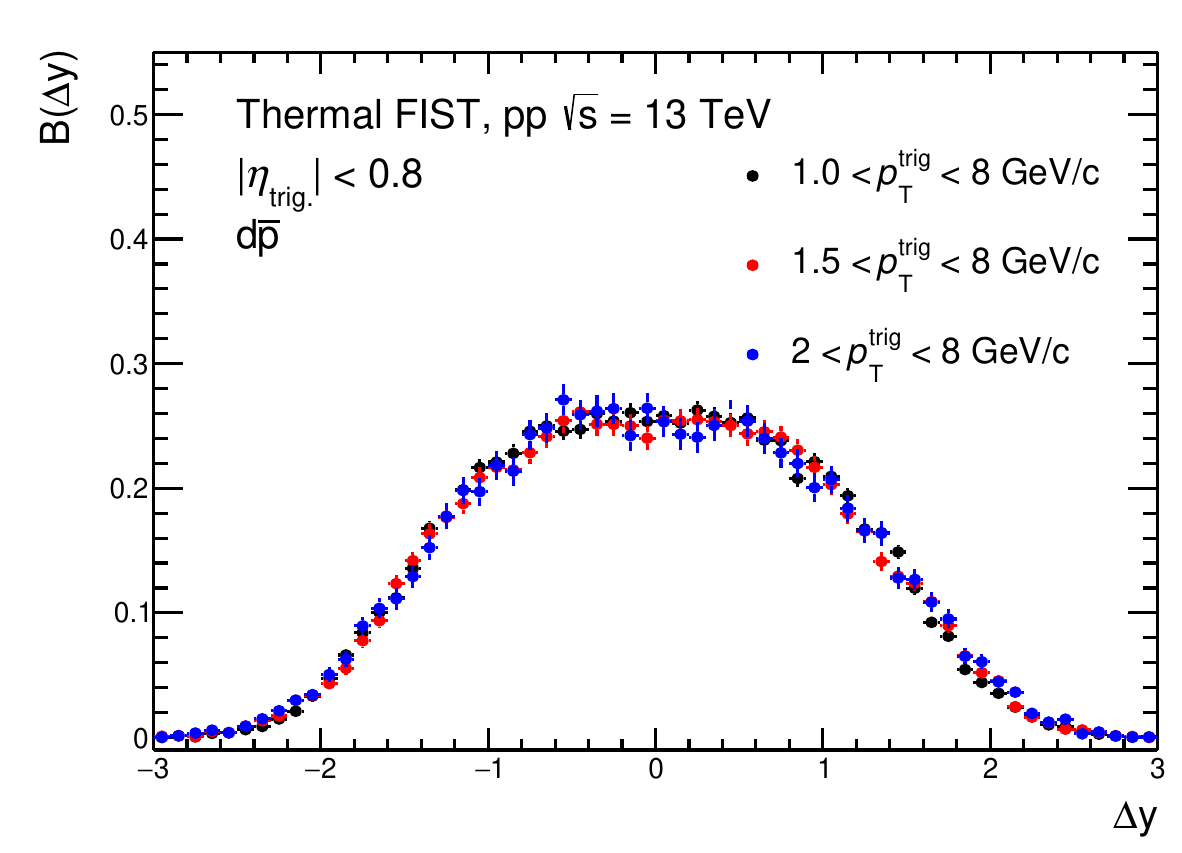}
\caption{Transverse momentum dependence of the  balance function of triggered protons and triggered deuterons from PYTHIA (left) and Thermal FIST (right) models.}
\label{fig:dp_pp_pTtrig}
\end{center}
\end{figure*}

\begin{figure*}[ht!]
\begin{center}
\includegraphics[scale=0.4]{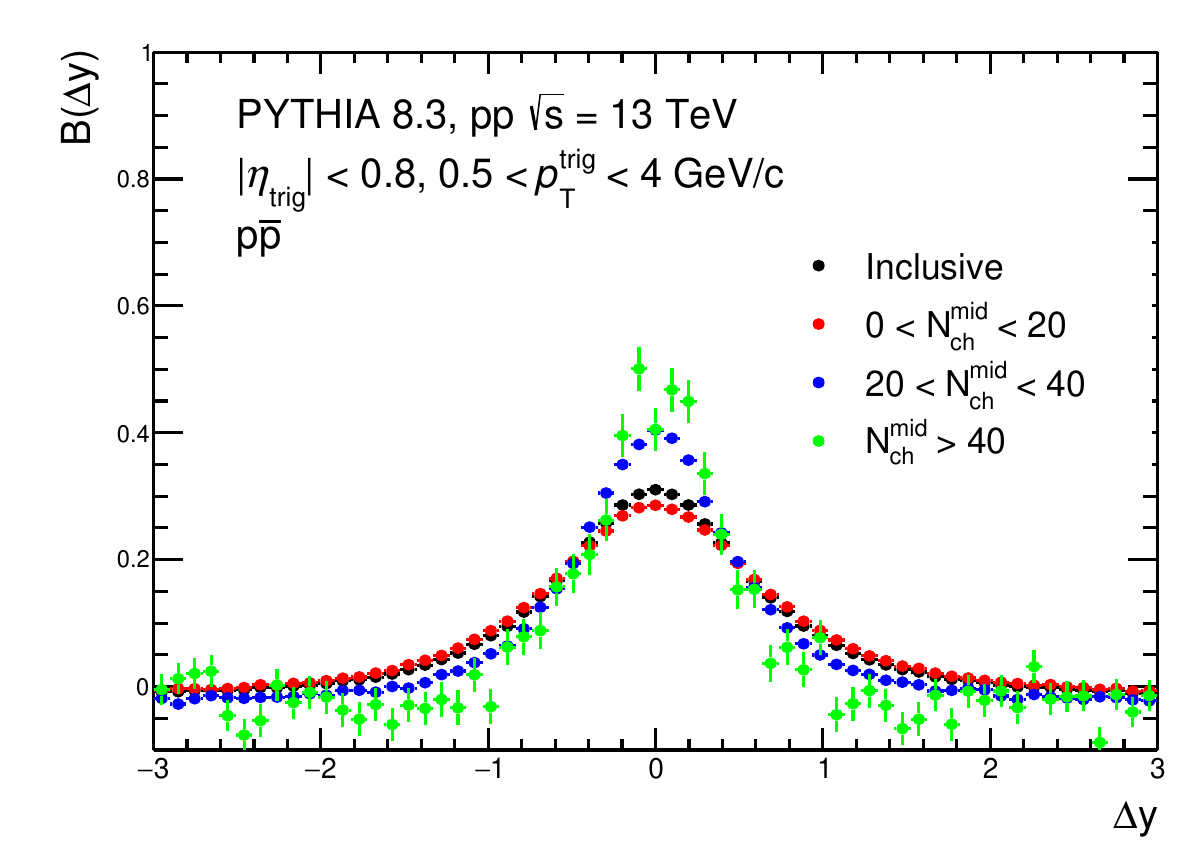}
\includegraphics[scale=0.4]{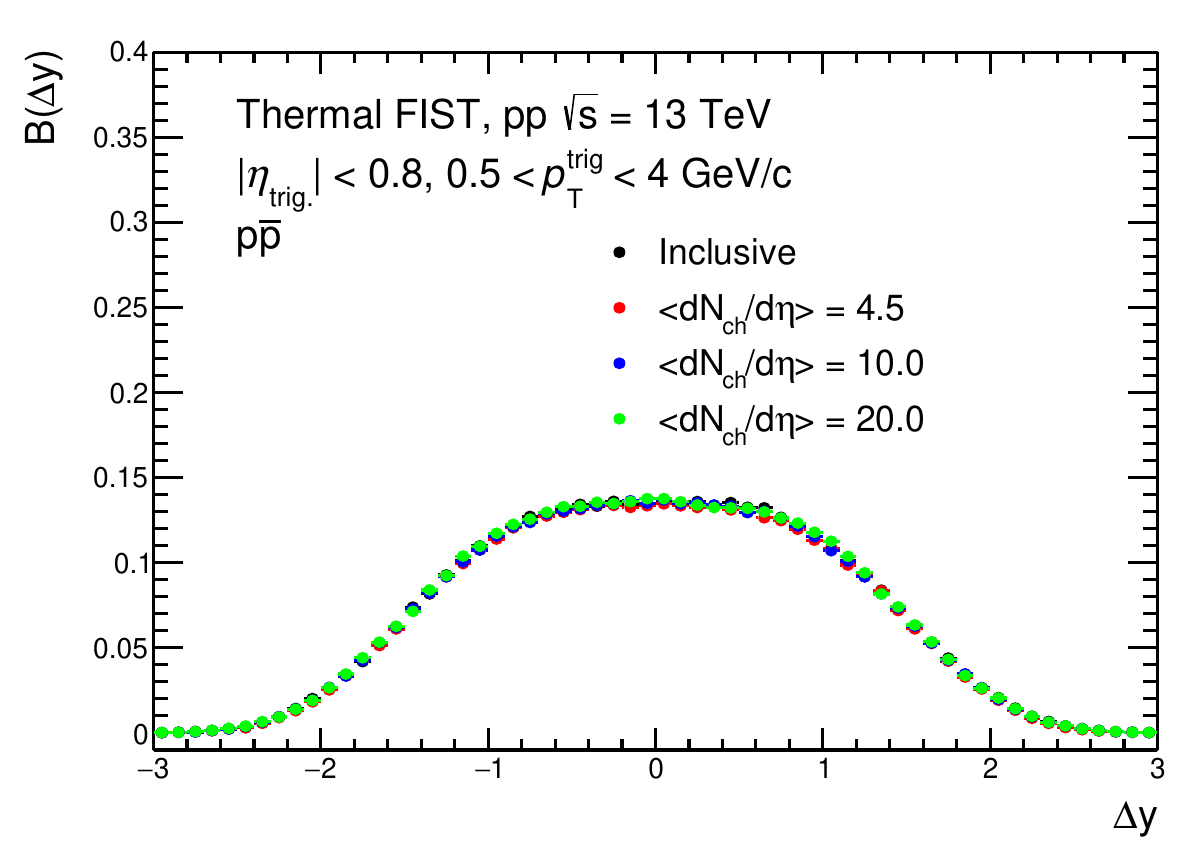}
\includegraphics[scale=0.4]{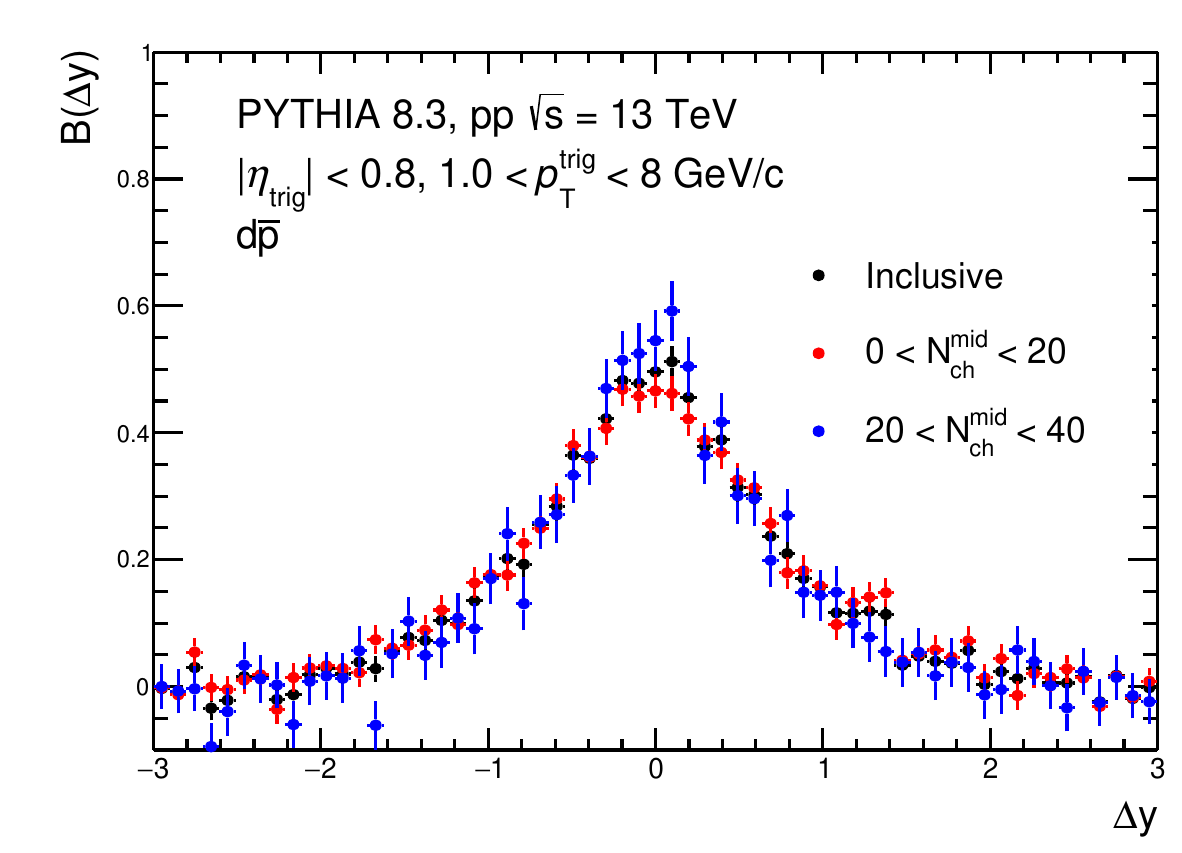}
\includegraphics[scale=0.4]{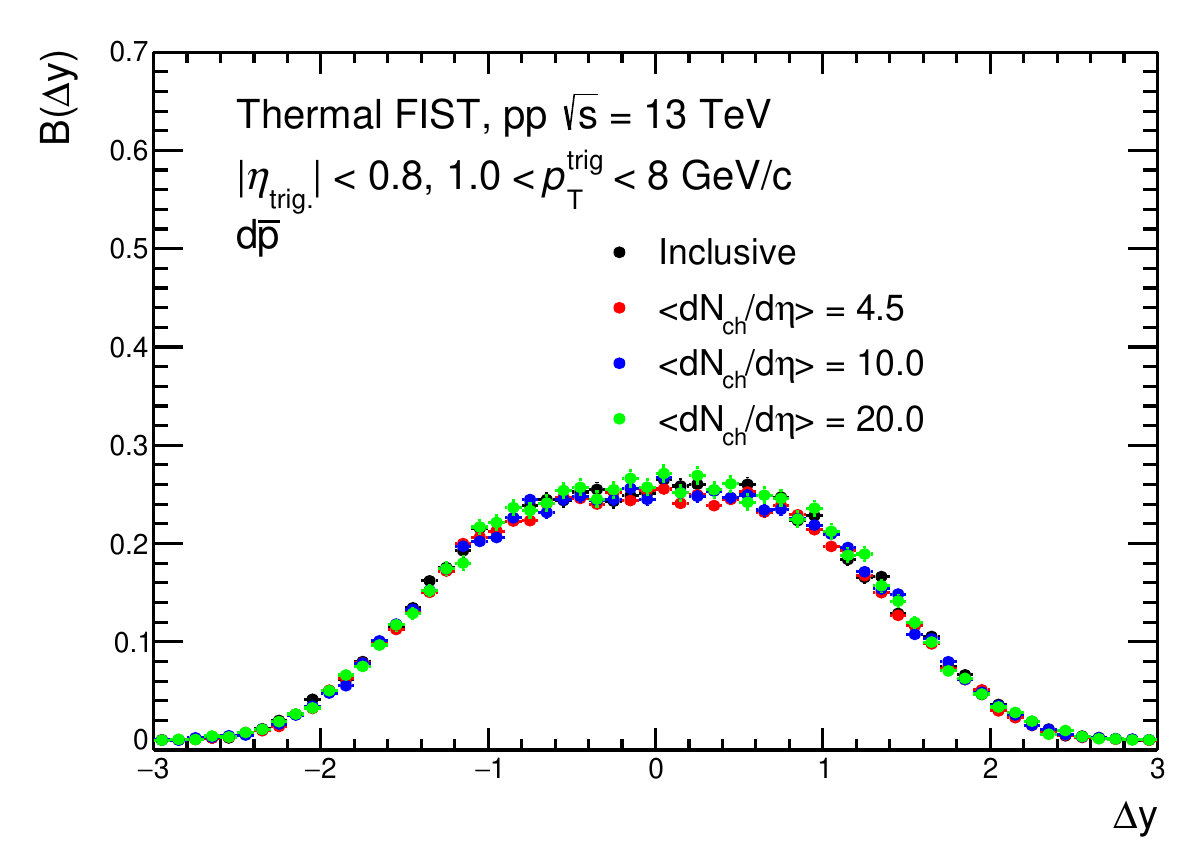}
\caption{Multiplicity dependence of the  balance function of triggered protons and triggered deuterons from PYTHIA (left) and Thermal FIST (right) models.}
\label{fig:dp_pp_pTtrigmult}
\end{center}
\end{figure*}

\begin{figure}
    \centering
\includegraphics[width=0.95\linewidth]{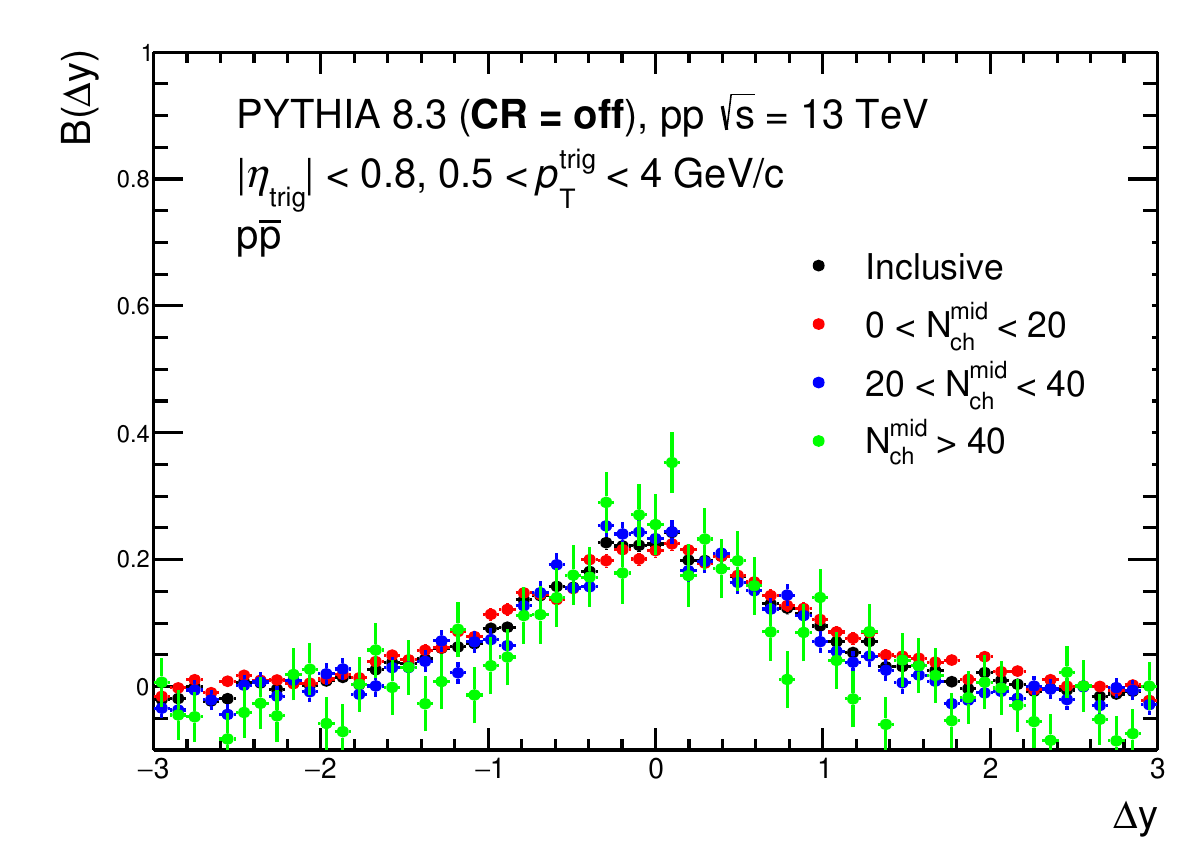}
    \caption{Multiplicity dependence of the balance function of triggered protons from PYTHIA with color reconnection  switched off}
\label{fig:dp_pp_pTtrigmultCRoff}
\end{figure}
What is interesting and slightly disappointing from these results is that, while the balance functions in Pythia and Thermal FIST are obviously different, there is the same relation between the deuteron and proton triggered results. The relation is rather intuitive for coalescence models as the balance by the deuterons is inherited from the nucleons. Our understanding is that for Thermal FIST the relation is satisfied because the shape of the balance function is the same for all hadrons because there are no microscopic correlations.

\begin{figure*}[ht!]
    \centering
\includegraphics[width=0.44\linewidth]{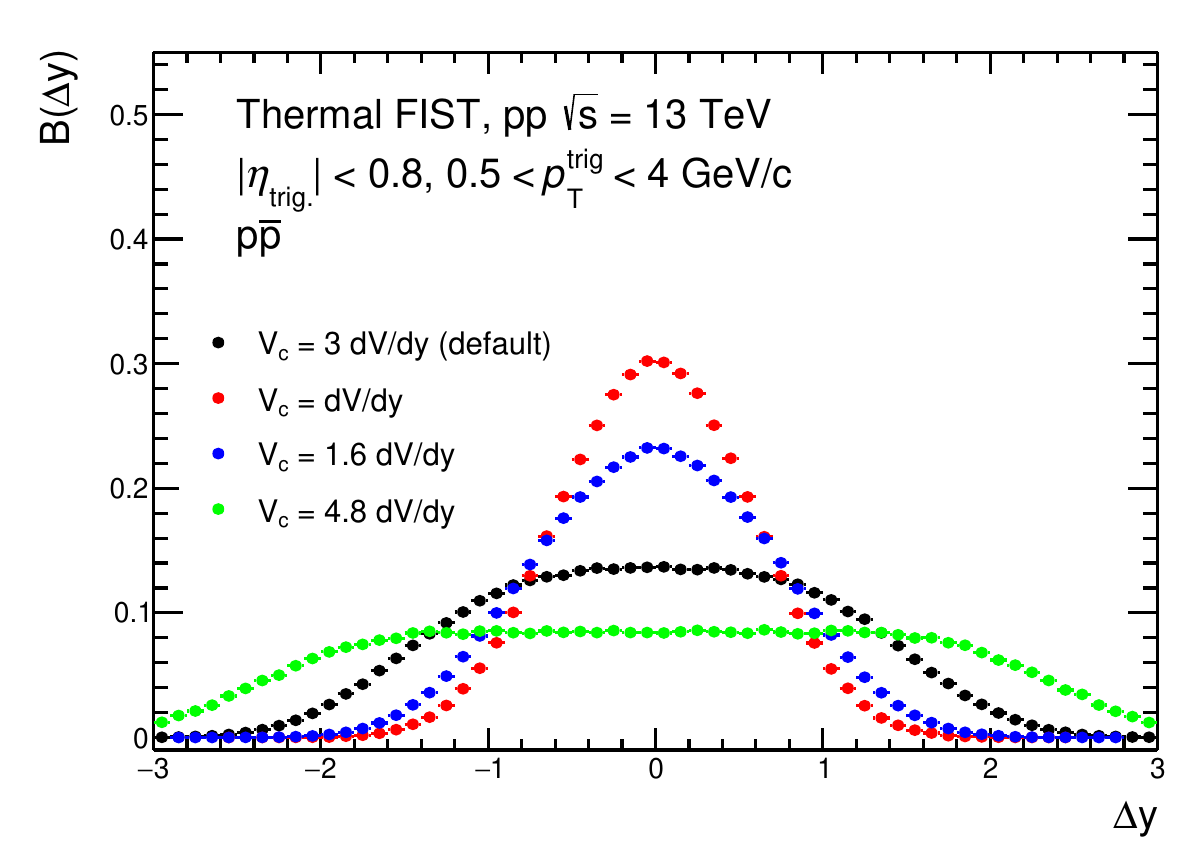}
\includegraphics[width=0.44\linewidth]{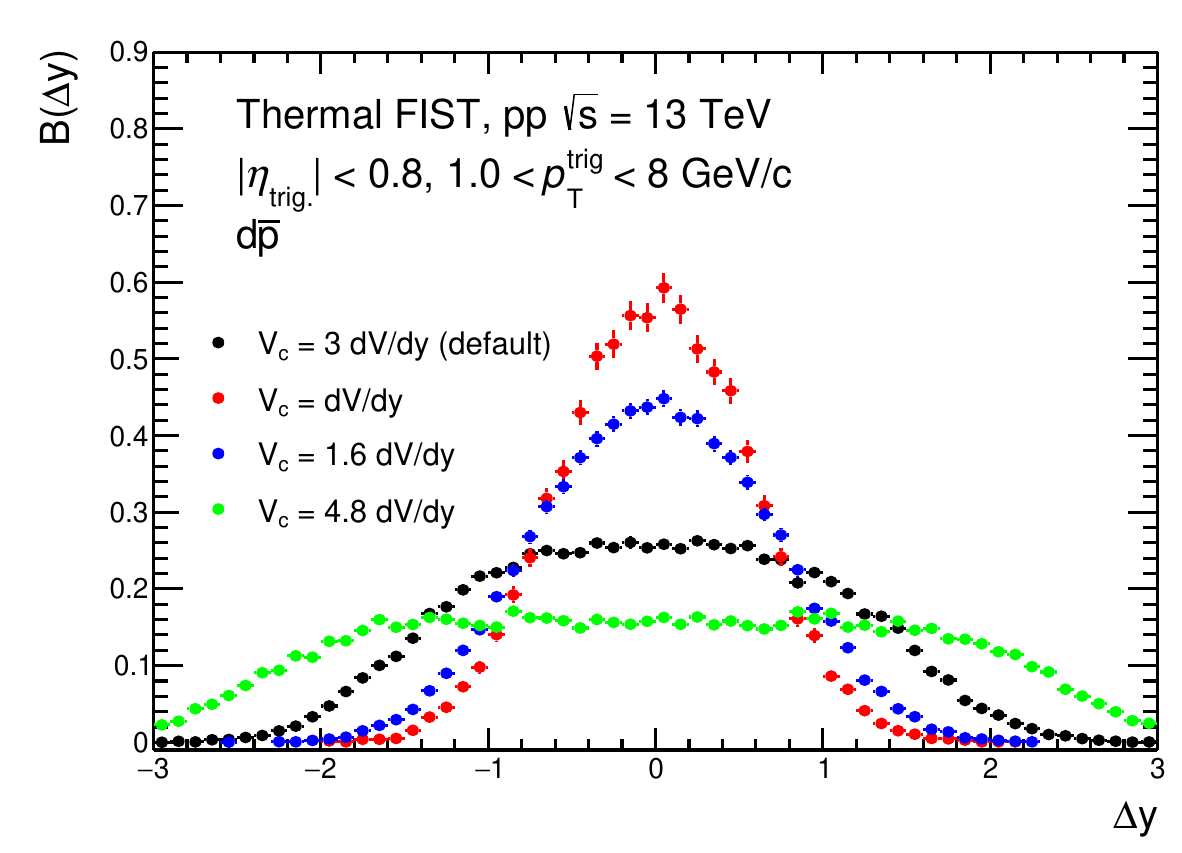}
    \caption{Correlation volume dependence of the  balance function of triggered protons and triggered deuterons from Thermal FIST.}
    \label{fig:corrvol}
\end{figure*}

\begin{figure*}[ht!]
    \centering
\includegraphics[width=0.44\linewidth]{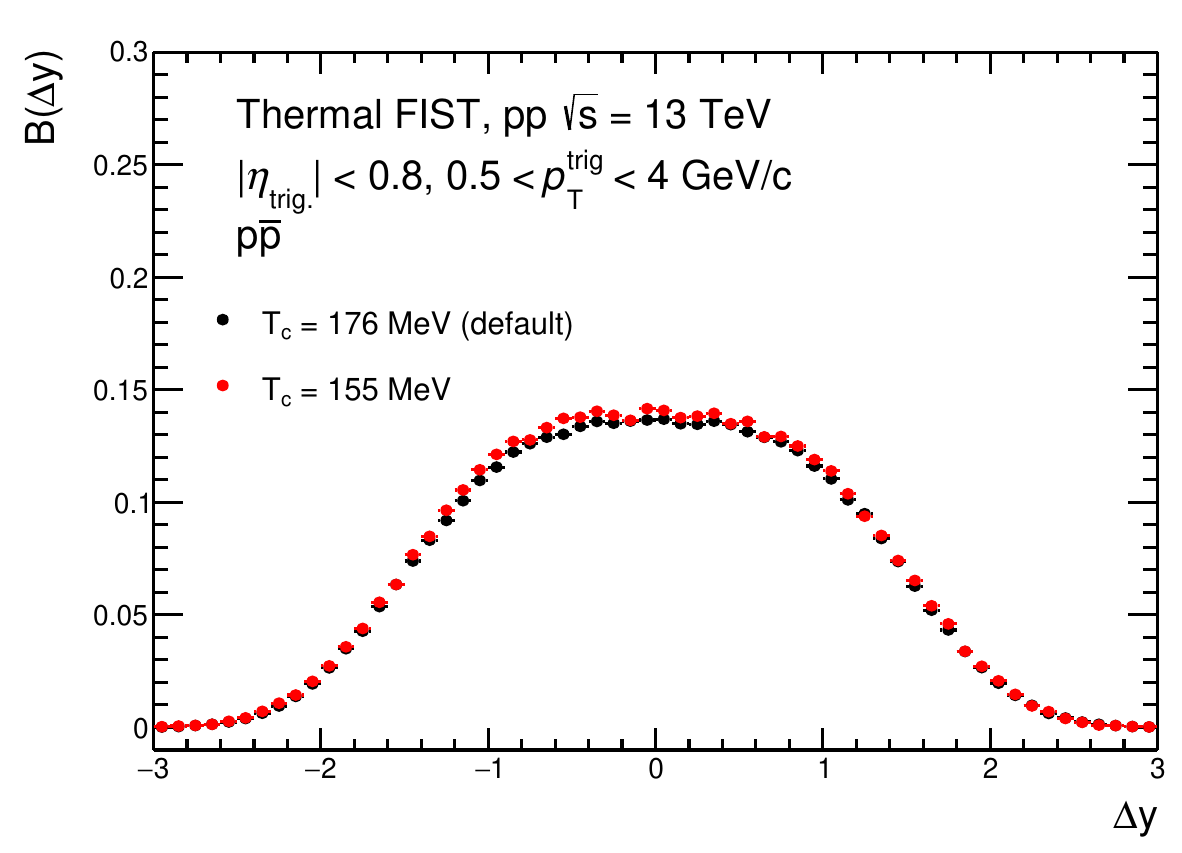}
\includegraphics[width=0.44\linewidth]{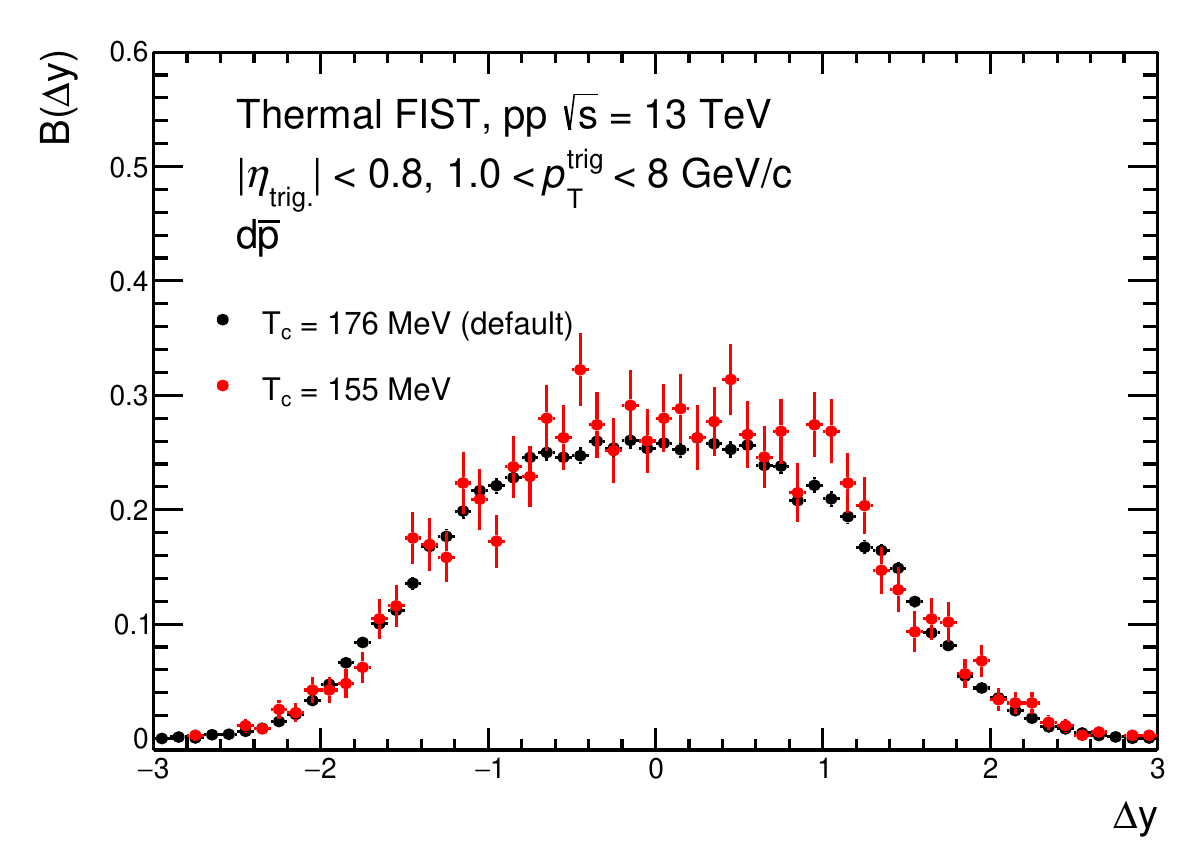}
    \caption{Temperature dependence of the  balance function of triggered protons and triggered deuterons from Thermal FIST.}
    \label{fig:temp}
\end{figure*}

This gives us an idea for how to differentiate between the models. In Pythia, the antibaryon that balance a proton is produced on the same string as the proton -- a color coherence effect. By varying the trigger $p_{\rm T}$ we can select the strings that are likely to be long and parallel in rapidity (low $p_{\rm T}^{\rm trig}$) or short in rapidity and extended in $p_{\rm T}$ (high $p_{\rm T}^{\rm trig}$).
We therefore study how the balance functions depend on the $p_{\rm T}$ of the trigger particle ($p_{\rm T}^{\rm trig}$) in the two models. Figure~\ref{fig:dp_pp_pTtrig} presents the results for triggered protons and deuterons across different $p_{\rm T}^{\rm trig}$ intervals. For the PYTHIA balance functions we observe the expected narrowing with $p_{\rm T}^{\rm trig}$ while the Thermal FIST balance functions shows no dependence on $p_{\rm T}^{\rm trig}$. This reflects that particles are produced independently in Thermal FIST with quantum number conservation only imposed globally. This striking difference in $p_{\rm T}^{\rm trig}$ dependence offers a promising handle to discriminate between the two production scenarios.

The result in Fig.~\ref{fig:dp_pp_pTtrig} is the main result of the paper and we want to discuss it a bit more in a broader context to highlight why we think it is so interesting. First, this result is generally applicable to all balance functions -- not only those affecting deuterons, protons, or baryons. Second, the results is really related to if particles and antiparticles are produced in the same process (the color coherence same string effect in PYTHIA) or if there is time for quantum numbers to decouple (as is the assumption in thermal models, e.g., Thermal FIST). Therefore, we think the difference is fundamental and directly related to the question of deconfinement.

To proceed with a more differential understanding, we next investigate the balance functions as a function of charged-particle multiplicity. Figure~\ref{fig:dp_pp_pTtrigmult} shows the multiplicity dependence of the balance function of triggered protons and triggered deuterons from PYTHIA and Thermal FIST models. For PYTHIA, the balance functions are obtained for three multiplicity classes, defined by the number of charged particles ($N_{\rm ch}^{\rm mid}$) at mid-rapidity ($|\eta|<0.8$) in the ranges 0–20, 20–40, and above 40. For Thermal FIST, the balance functions are calculated for $\langle {\rm d}N_{\rm ch}/d\eta\rangle$ values of 4.5, 10, and 20, corresponding to the (0–1)\%, (10–20)\%, and (40–50)\% multiplicity classes determined by the V0 detectors of the ALICE Collaboration~\cite{ALICE:2020swj}. In PYTHIA, a modest multiplicity dependence is observed for the balance functions of both proton- and deuteron-triggered events. In contrast, in Thermal-FIST, no multiplicity dependence is observed. The effect in PYTHIA can be attributed to the increase of the mean transverse momentum with multiplicity, which in PYTHIA primarily arises from color reconnection (CR). Once CR is switched off, as illustrated in Fig.~\ref{fig:dp_pp_pTtrigmultCRoff}, the multiplicity dependence of balance functions with triggered protons disappears. 

To investigate the origin of the broader balance functions in Thermal-FIST, we vary the correlation volume, $V_{\rm c}$, using three values: $V_{\rm c} = \dvdy$, $V_{\rm c} = 1.6\,\dvdy$, and $V_{\rm c} = 4.8\,\dvdy$. The choice of factors 1.6 and 4.8 are inspired by a recent ALICE study~\cite{ALICE:2022xiu} on antideuteron number fluctuations in Pb--Pb collisions. Figure~\ref{fig:corrvol} shows the correlation-volume dependence of the balance functions for triggered protons and deuterons. When the correlation volume is reduced from $4.8{\rm d}V/{\rm d}y$ to ${\rm d}V/{\rm d}y$, the balance functions become narrower, indicating that more balancing pairs are found at mid-rapidity. This highlights the sensitivity of the balance functions to the correlation length. Note that the rapidity shape depends on how one defines the correlation volume and is one of the parameters that are not constrained in Thermal-FIST~\cite{Vovchenko:2024pvk}. 

To explore the temperature dependence in Thermal FIST, we perform a test by lowering the temperature from the default $T_{c} = 176$ to $155$ MeV. As shown in Fig.\ref{fig:temp}, this change has no visible effect on the balance functions for either protons or deuterons. A similar observation was reported for the $\Xi$–K balance function in Ref.~\cite{Bierlich:2025pkg}. 

\section{Conclusion}
\label{conclusion}
In this work, we investigated deuteron- and proton-triggered balance functions in pp collisions to test if we can be sensitive to the microscopic mechanisms underlying their formation.

Using the coalescence model implemented in PYTHIA and the statistical model in the Thermal FIST package, we presented balance functions for triggered protons and deuterons. PYTHIA and Thermal FIST exhibit visibly different balance function shapes for both protons and deuterons. The broader balance functions observed in Thermal FIST are directly related to the choice of correlation volume. In both models, the balance function of triggered protons approximately matches that of triggered deuterons scaled by a factor of 1/2, consistent with the intuitive expectation that a deuteron, composed of two nucleons, should have a balance function approximately twice that of a proton.

In addition, we find that deuteron–pion balances vanish identically in both models. This can be understood from baryon number and charge conservation together with isospin symmetry, which ensures that the contributions from proton–$\pi^-$ and neutron–$\pi^+$ channels cancel exactly for a deuteron. 

A moderate multiplicity dependence of the balance functions is observed in PYTHIA, while Thermal FIST shows no such dependence. A particularly striking difference emerges in the transverse momentum dependence of the balance functions: PYTHIA exhibits a clear narrowing with increasing trigger $p_{\rm T}$, whereas Thermal FIST shows no such dependence. This observable therefore offers a promising discriminator between the two production mechanisms scenarios and motivates using the data already collected during Run 3 of the LHC. Such measurements would provide important insight on the microscopic production mechanism of nuclei in high-energy collisions and could even be used to test hadron production mechanisms in general.


\section{Acknowledgment}
The authors acknowledge the funding received from the European Union’s Horizon Europe research and innovation programme under the Marie Skłodowska-Curie grant agreement No. 101149298. 

\bibliography{References}

\begin{thebibliography}{42}%
\makeatletter
\providecommand \@ifxundefined [1]{%
 \@ifx{#1\undefined}
}%
\providecommand \@ifnum [1]{%
 \ifnum #1\expandafter \@firstoftwo
 \else \expandafter \@secondoftwo
 \fi
}%
\providecommand \@ifx [1]{%
 \ifx #1\expandafter \@firstoftwo
 \else \expandafter \@secondoftwo
 \fi
}%
\providecommand \natexlab [1]{#1}%
\providecommand \enquote  [1]{``#1''}%
\providecommand \bibnamefont  [1]{#1}%
\providecommand \bibfnamefont [1]{#1}%
\providecommand \citenamefont [1]{#1}%
\providecommand \href@noop [0]{\@secondoftwo}%
\providecommand \href [0]{\begingroup \@sanitize@url \@href}%
\providecommand \@href[1]{\@@startlink{#1}\@@href}%
\providecommand \@@href[1]{\endgroup#1\@@endlink}%
\providecommand \@sanitize@url [0]{\catcode `\\12\catcode `\$12\catcode
  `\&12\catcode `\#12\catcode `\^12\catcode `\_12\catcode `\%12\relax}%
\providecommand \@@startlink[1]{}%
\providecommand \@@endlink[0]{}%
\providecommand \url  [0]{\begingroup\@sanitize@url \@url }%
\providecommand \@url [1]{\endgroup\@href {#1}{\urlprefix }}%
\providecommand \urlprefix  [0]{URL }%
\providecommand \Eprint [0]{\href }%
\providecommand \doibase [0]{https://doi.org/}%
\providecommand \selectlanguage [0]{\@gobble}%
\providecommand \bibinfo  [0]{\@secondoftwo}%
\providecommand \bibfield  [0]{\@secondoftwo}%
\providecommand \translation [1]{[#1]}%
\providecommand \BibitemOpen [0]{}%
\providecommand \bibitemStop [0]{}%
\providecommand \bibitemNoStop [0]{.\EOS\space}%
\providecommand \EOS [0]{\spacefactor3000\relax}%
\providecommand \BibitemShut  [1]{\csname bibitem#1\endcsname}%
\let\auto@bib@innerbib\@empty
\bibitem [{\citenamefont {Adam}\ \emph {et~al.}(2016)\citenamefont {Adam} \emph
  {et~al.}}]{nuclei_pp_PbPb}%
  \BibitemOpen
  \bibfield  {author} {\bibinfo {author} {\bibfnamefont {J.}~\bibnamefont
  {Adam}} \emph {et~al.} (\bibinfo {collaboration} {ALICE}),\ }\href
  {https://doi.org/10.1103/PhysRevC.93.024917} {\bibfield  {journal} {\bibinfo
  {journal} {Phys. Rev. C}\ }\textbf {\bibinfo {volume} {93}},\ \bibinfo
  {pages} {024917} (\bibinfo {year} {2016})},\ \Eprint
  {https://arxiv.org/abs/1506.08951} {arXiv:1506.08951 [nucl-ex]} \BibitemShut
  {NoStop}%
\bibitem [{\citenamefont {Acharya}\ \emph {et~al.}(2018)\citenamefont {Acharya}
  \emph {et~al.}}]{nuclei_pp}%
  \BibitemOpen
  \bibfield  {author} {\bibinfo {author} {\bibfnamefont {S.}~\bibnamefont
  {Acharya}} \emph {et~al.} (\bibinfo {collaboration} {ALICE}),\ }\href
  {https://doi.org/10.1103/PhysRevC.97.024615} {\bibfield  {journal} {\bibinfo
  {journal} {Phys. Rev.}\ }\textbf {\bibinfo {volume} {C97}},\ \bibinfo {pages}
  {024615} (\bibinfo {year} {2018})},\ \Eprint
  {https://arxiv.org/abs/1709.08522} {arXiv:1709.08522 [nucl-ex]} \BibitemShut
  {NoStop}%
\bibitem [{\citenamefont {Acharya}\ \emph {et~al.}(2019)\citenamefont {Acharya}
  \emph {et~al.}}]{deuteron_pp_7TeV}%
  \BibitemOpen
  \bibfield  {author} {\bibinfo {author} {\bibfnamefont {S.}~\bibnamefont
  {Acharya}} \emph {et~al.} (\bibinfo {collaboration} {ALICE}),\ }\href
  {https://doi.org/10.1016/j.physletb.2019.05.028} {\bibfield  {journal}
  {\bibinfo  {journal} {Phys. Lett.}\ }\textbf {\bibinfo {volume} {B794}},\
  \bibinfo {pages} {50} (\bibinfo {year} {2019})},\ \Eprint
  {https://arxiv.org/abs/1902.09290} {arXiv:1902.09290 [nucl-ex]} \BibitemShut
  {NoStop}%
\bibitem [{\citenamefont {Acharya}\ \emph
  {et~al.}(2020{\natexlab{a}})\citenamefont {Acharya} \emph
  {et~al.}}]{deuteron_pPbALICE}%
  \BibitemOpen
  \bibfield  {author} {\bibinfo {author} {\bibfnamefont {S.}~\bibnamefont
  {Acharya}} \emph {et~al.} (\bibinfo {collaboration} {ALICE}),\ }\href
  {https://doi.org/10.1016/j.physletb.2019.135043} {\bibfield  {journal}
  {\bibinfo  {journal} {Phys. Lett. B}\ }\textbf {\bibinfo {volume} {800}},\
  \bibinfo {pages} {135043} (\bibinfo {year} {2020}{\natexlab{a}})},\ \Eprint
  {https://arxiv.org/abs/1906.03136} {arXiv:1906.03136 [nucl-ex]} \BibitemShut
  {NoStop}%
\bibitem [{\citenamefont {Acharya}\ \emph
  {et~al.}(2020{\natexlab{b}})\citenamefont {Acharya} \emph
  {et~al.}}]{3He_pPb}%
  \BibitemOpen
  \bibfield  {author} {\bibinfo {author} {\bibfnamefont {S.}~\bibnamefont
  {Acharya}} \emph {et~al.} (\bibinfo {collaboration} {ALICE}),\ }\href
  {https://doi.org/10.1103/PhysRevC.101.044906} {\bibfield  {journal} {\bibinfo
   {journal} {Phys. Rev.}\ }\textbf {\bibinfo {volume} {C101}},\ \bibinfo
  {pages} {044906} (\bibinfo {year} {2020}{\natexlab{b}})},\ \Eprint
  {https://arxiv.org/abs/1910.14401} {arXiv:1910.14401 [nucl-ex]} \BibitemShut
  {NoStop}%
\bibitem [{\citenamefont {Acharya}\ \emph
  {et~al.}(2020{\natexlab{c}})\citenamefont {Acharya} \emph
  {et~al.}}]{deuteron_pp_13TeV}%
  \BibitemOpen
  \bibfield  {author} {\bibinfo {author} {\bibfnamefont {S.}~\bibnamefont
  {Acharya}} \emph {et~al.} (\bibinfo {collaboration} {ALICE}),\ }\href
  {https://doi.org/10.1140/epjc/s10052-020-8256-4} {\bibfield  {journal}
  {\bibinfo  {journal} {Eur. Phys. J. C}\ }\textbf {\bibinfo {volume} {80}},\
  \bibinfo {pages} {889} (\bibinfo {year} {2020}{\natexlab{c}})},\ \Eprint
  {https://arxiv.org/abs/2003.03184} {arXiv:2003.03184 [nucl-ex]} \BibitemShut
  {NoStop}%
\bibitem [{\citenamefont {Acharya}\ \emph
  {et~al.}(2022{\natexlab{a}})\citenamefont {Acharya} \emph
  {et~al.}}]{nuclei_pp_5TeV}%
  \BibitemOpen
  \bibfield  {author} {\bibinfo {author} {\bibfnamefont {S.}~\bibnamefont
  {Acharya}} \emph {et~al.} (\bibinfo {collaboration} {ALICE}),\ }\href
  {https://doi.org/10.1140/epjc/s10052-022-10241-z} {\bibfield  {journal}
  {\bibinfo  {journal} {Eur. Phys. J. C}\ }\textbf {\bibinfo {volume} {82}},\
  \bibinfo {pages} {289} (\bibinfo {year} {2022}{\natexlab{a}})},\ \Eprint
  {https://arxiv.org/abs/2112.00610} {arXiv:2112.00610 [nucl-ex]} \BibitemShut
  {NoStop}%
\bibitem [{\citenamefont {Acharya}\ \emph
  {et~al.}(2022{\natexlab{b}})\citenamefont {Acharya} \emph
  {et~al.}}]{nuclei_pp_13TeV_HM}%
  \BibitemOpen
  \bibfield  {author} {\bibinfo {author} {\bibfnamefont {S.}~\bibnamefont
  {Acharya}} \emph {et~al.} (\bibinfo {collaboration} {ALICE}),\ }\href
  {https://doi.org/10.1007/JHEP01(2022)106} {\bibfield  {journal} {\bibinfo
  {journal} {JHEP}\ }\textbf {\bibinfo {volume} {01}},\ \bibinfo {pages}
  {106}},\ \Eprint {https://arxiv.org/abs/2109.13026} {arXiv:2109.13026
  [nucl-ex]} \BibitemShut {NoStop}%
\bibitem [{\citenamefont {Bierlich}\ \emph {et~al.}(2022)\citenamefont
  {Bierlich} \emph {et~al.}}]{Bierlich:2022pfr}%
  \BibitemOpen
  \bibfield  {author} {\bibinfo {author} {\bibfnamefont {C.}~\bibnamefont
  {Bierlich}} \emph {et~al.},\ }\href
  {https://doi.org/10.21468/SciPostPhysCodeb.8} {\bibfield  {journal} {\bibinfo
   {journal} {SciPost Phys. Codeb.}\ }\textbf {\bibinfo {volume} {2022}},\
  \bibinfo {pages} {8} (\bibinfo {year} {2022})},\ \Eprint
  {https://arxiv.org/abs/2203.11601} {arXiv:2203.11601 [hep-ph]} \BibitemShut
  {NoStop}%
\bibitem [{\citenamefont {Pierog}\ \emph {et~al.}(2015)\citenamefont {Pierog},
  \citenamefont {Karpenko}, \citenamefont {Katzy}, \citenamefont {Yatsenko},\
  and\ \citenamefont {Werner}}]{Pierog:2013ria}%
  \BibitemOpen
  \bibfield  {author} {\bibinfo {author} {\bibfnamefont {T.}~\bibnamefont
  {Pierog}}, \bibinfo {author} {\bibfnamefont {I.}~\bibnamefont {Karpenko}},
  \bibinfo {author} {\bibfnamefont {J.~M.}\ \bibnamefont {Katzy}}, \bibinfo
  {author} {\bibfnamefont {E.}~\bibnamefont {Yatsenko}},\ and\ \bibinfo
  {author} {\bibfnamefont {K.}~\bibnamefont {Werner}},\ }\href
  {https://doi.org/10.1103/PhysRevC.92.034906} {\bibfield  {journal} {\bibinfo
  {journal} {Phys. Rev. C}\ }\textbf {\bibinfo {volume} {92}},\ \bibinfo
  {pages} {034906} (\bibinfo {year} {2015})},\ \Eprint
  {https://arxiv.org/abs/1306.0121} {arXiv:1306.0121 [hep-ph]} \BibitemShut
  {NoStop}%
\bibitem [{\citenamefont {Korsmeier}\ \emph {et~al.}(2018)\citenamefont
  {Korsmeier}, \citenamefont {Donato},\ and\ \citenamefont
  {Fornengo}}]{Korsmeier:2017xzj}%
  \BibitemOpen
  \bibfield  {author} {\bibinfo {author} {\bibfnamefont {M.}~\bibnamefont
  {Korsmeier}}, \bibinfo {author} {\bibfnamefont {F.}~\bibnamefont {Donato}},\
  and\ \bibinfo {author} {\bibfnamefont {N.}~\bibnamefont {Fornengo}},\ }\href
  {https://doi.org/10.1103/PhysRevD.97.103011} {\bibfield  {journal} {\bibinfo
  {journal} {Phys.Rev.D}\ }\textbf {\bibinfo {volume} {97}},\ \bibinfo {pages}
  {103011} (\bibinfo {year} {2018})},\ \Eprint
  {https://arxiv.org/abs/1711.08465} {arXiv:1711.08465 [astro-ph.HE]}
  \BibitemShut {NoStop}%
\bibitem [{\citenamefont {Blum}\ \emph {et~al.}(2017)\citenamefont {Blum},
  \citenamefont {Ng}, \citenamefont {Sato},\ and\ \citenamefont
  {Takimoto}}]{Blum:2017qnn}%
  \BibitemOpen
  \bibfield  {author} {\bibinfo {author} {\bibfnamefont {K.}~\bibnamefont
  {Blum}}, \bibinfo {author} {\bibfnamefont {K.~C.~Y.}\ \bibnamefont {Ng}},
  \bibinfo {author} {\bibfnamefont {R.}~\bibnamefont {Sato}},\ and\ \bibinfo
  {author} {\bibfnamefont {M.}~\bibnamefont {Takimoto}},\ }\href
  {https://doi.org/10.1103/PhysRevD.96.103021} {\bibfield  {journal} {\bibinfo
  {journal} {Phys. Rev. D}\ }\textbf {\bibinfo {volume} {96}},\ \bibinfo
  {pages} {103021} (\bibinfo {year} {2017})},\ \Eprint
  {https://arxiv.org/abs/1704.05431} {arXiv:1704.05431 [astro-ph.HE]}
  \BibitemShut {NoStop}%
\bibitem [{\citenamefont {Kachelrie\ss{}}\ \emph {et~al.}(2020)\citenamefont
  {Kachelrie\ss{}}, \citenamefont {Ostapchenko},\ and\ \citenamefont
  {Tjemsland}}]{Kachelriess:2020uoh}%
  \BibitemOpen
  \bibfield  {author} {\bibinfo {author} {\bibfnamefont {M.}~\bibnamefont
  {Kachelrie\ss{}}}, \bibinfo {author} {\bibfnamefont {S.}~\bibnamefont
  {Ostapchenko}},\ and\ \bibinfo {author} {\bibfnamefont {J.}~\bibnamefont
  {Tjemsland}},\ }\href {https://doi.org/10.1088/1475-7516/2020/08/048}
  {\bibfield  {journal} {\bibinfo  {journal} {JCAP}\ }\textbf {\bibinfo
  {volume} {08}},\ \bibinfo {pages} {048}},\ \Eprint
  {https://arxiv.org/abs/2002.10481} {arXiv:2002.10481 [hep-ph]} \BibitemShut
  {NoStop}%
\bibitem [{\citenamefont {von Doetinchem}\ \emph {et~al.}(2020)\citenamefont
  {von Doetinchem} \emph {et~al.}}]{vonDoetinchem:2020vbj}%
  \BibitemOpen
  \bibfield  {author} {\bibinfo {author} {\bibfnamefont {P.}~\bibnamefont {von
  Doetinchem}} \emph {et~al.},\ }\href
  {https://doi.org/10.1088/1475-7516/2020/08/035} {\bibfield  {journal}
  {\bibinfo  {journal} {JCAP}\ }\textbf {\bibinfo {volume} {08}},\ \bibinfo
  {pages} {035}},\ \Eprint {https://arxiv.org/abs/2002.04163} {arXiv:2002.04163
  [astro-ph.HE]} \BibitemShut {NoStop}%
\bibitem [{\citenamefont {\ifmmode \check{S}\else \v{S}\fi{}erk\ifmmode
  \check{s}\else \v{s}\fi{}nyt\ifmmode~\dot{e}\else \.{e}\fi{}}\ \emph
  {et~al.}(2022)\citenamefont {\ifmmode \check{S}\else \v{S}\fi{}erk\ifmmode
  \check{s}\else \v{s}\fi{}nyt\ifmmode~\dot{e}\else \.{e}\fi{}}, \citenamefont
  {K\"onigstorfer}, \citenamefont {von Doetinchem}, \citenamefont {Fabbietti},
  \citenamefont {Gomez-Coral}, \citenamefont {Herms}, \citenamefont {Ibarra},
  \citenamefont {P\"oschl}, \citenamefont {Shukla}, \citenamefont {Strong},\
  and\ \citenamefont {Vorobyev}}]{PhysRevD.105.083021}%
  \BibitemOpen
  \bibfield  {author} {\bibinfo {author} {\bibfnamefont {L.}~\bibnamefont
  {\ifmmode \check{S}\else \v{S}\fi{}erk\ifmmode \check{s}\else
  \v{s}\fi{}nyt\ifmmode~\dot{e}\else \.{e}\fi{}}}, \bibinfo {author}
  {\bibfnamefont {S.}~\bibnamefont {K\"onigstorfer}}, \bibinfo {author}
  {\bibfnamefont {P.}~\bibnamefont {von Doetinchem}}, \bibinfo {author}
  {\bibfnamefont {L.}~\bibnamefont {Fabbietti}}, \bibinfo {author}
  {\bibfnamefont {D.~M.}\ \bibnamefont {Gomez-Coral}}, \bibinfo {author}
  {\bibfnamefont {J.}~\bibnamefont {Herms}}, \bibinfo {author} {\bibfnamefont
  {A.}~\bibnamefont {Ibarra}}, \bibinfo {author} {\bibfnamefont
  {T.}~\bibnamefont {P\"oschl}}, \bibinfo {author} {\bibfnamefont
  {A.}~\bibnamefont {Shukla}}, \bibinfo {author} {\bibfnamefont
  {A.}~\bibnamefont {Strong}},\ and\ \bibinfo {author} {\bibfnamefont
  {I.}~\bibnamefont {Vorobyev}},\ }\href
  {https://doi.org/10.1103/PhysRevD.105.083021} {\bibfield  {journal} {\bibinfo
   {journal} {Phys. Rev. D}\ }\textbf {\bibinfo {volume} {105}},\ \bibinfo
  {pages} {083021} (\bibinfo {year} {2022})}\BibitemShut {NoStop}%
\bibitem [{\citenamefont {Sato}\ and\ \citenamefont
  {Yazaki}(1981)}]{Sato:1981ez}%
  \BibitemOpen
  \bibfield  {author} {\bibinfo {author} {\bibfnamefont {H.}~\bibnamefont
  {Sato}}\ and\ \bibinfo {author} {\bibfnamefont {K.}~\bibnamefont {Yazaki}},\
  }\href {https://doi.org/10.1016/0370-2693(81)90976-X} {\bibfield  {journal}
  {\bibinfo  {journal} {Phys. Lett.}\ }\textbf {\bibinfo {volume} {B98}},\
  \bibinfo {pages} {153} (\bibinfo {year} {1981})}\BibitemShut {NoStop}%
\bibitem [{\citenamefont {Nagle}\ \emph {et~al.}(1996)\citenamefont {Nagle},
  \citenamefont {Kumar}, \citenamefont {Kusnezov}, \citenamefont {Sorge},\ and\
  \citenamefont {Mattiello}}]{Nagle:1996vp}%
  \BibitemOpen
  \bibfield  {author} {\bibinfo {author} {\bibfnamefont {J.~L.}\ \bibnamefont
  {Nagle}}, \bibinfo {author} {\bibfnamefont {B.~S.}\ \bibnamefont {Kumar}},
  \bibinfo {author} {\bibfnamefont {D.}~\bibnamefont {Kusnezov}}, \bibinfo
  {author} {\bibfnamefont {H.}~\bibnamefont {Sorge}},\ and\ \bibinfo {author}
  {\bibfnamefont {R.}~\bibnamefont {Mattiello}},\ }\href
  {https://doi.org/10.1103/PhysRevC.53.367} {\bibfield  {journal} {\bibinfo
  {journal} {Phys. Rev.}\ }\textbf {\bibinfo {volume} {C53}},\ \bibinfo {pages}
  {367} (\bibinfo {year} {1996})}\BibitemShut {NoStop}%
\bibitem [{\citenamefont {Scheibl}\ and\ \citenamefont
  {Heinz}(1999)}]{Scheibl:1998tk}%
  \BibitemOpen
  \bibfield  {author} {\bibinfo {author} {\bibfnamefont {R.}~\bibnamefont
  {Scheibl}}\ and\ \bibinfo {author} {\bibfnamefont {U.~W.}\ \bibnamefont
  {Heinz}},\ }\href {https://doi.org/10.1103/PhysRevC.59.1585} {\bibfield
  {journal} {\bibinfo  {journal} {Phys. Rev.}\ }\textbf {\bibinfo {volume}
  {C59}},\ \bibinfo {pages} {1585} (\bibinfo {year} {1999})},\ \Eprint
  {https://arxiv.org/abs/nucl-th/9809092} {arXiv:nucl-th/9809092 [nucl-th]}
  \BibitemShut {NoStop}%
\bibitem [{\citenamefont {Blum}\ and\ \citenamefont
  {Takimoto}(2019)}]{Blum:2019suo}%
  \BibitemOpen
  \bibfield  {author} {\bibinfo {author} {\bibfnamefont {K.}~\bibnamefont
  {Blum}}\ and\ \bibinfo {author} {\bibfnamefont {M.}~\bibnamefont
  {Takimoto}},\ }\href {https://doi.org/10.1103/PhysRevC.99.044913} {\bibfield
  {journal} {\bibinfo  {journal} {Phys. Rev.}\ }\textbf {\bibinfo {volume}
  {C99}},\ \bibinfo {pages} {044913} (\bibinfo {year} {2019})},\ \Eprint
  {https://arxiv.org/abs/1901.07088} {arXiv:1901.07088 [nucl-th]} \BibitemShut
  {NoStop}%
\bibitem [{\citenamefont {Mr\'owczy\'nski}\ and\ \citenamefont
  {S\l{}o\'n}(2020)}]{Mrowczynski:2019yrr}%
  \BibitemOpen
  \bibfield  {author} {\bibinfo {author} {\bibfnamefont {S.}~\bibnamefont
  {Mr\'owczy\'nski}}\ and\ \bibinfo {author} {\bibfnamefont {P.}~\bibnamefont
  {S\l{}o\'n}},\ }\href {https://doi.org/10.5506/APhysPolB.51.1739} {\bibfield
  {journal} {\bibinfo  {journal} {Acta Phys. Polon. B}\ }\textbf {\bibinfo
  {volume} {51}},\ \bibinfo {pages} {1739} (\bibinfo {year} {2020})},\ \Eprint
  {https://arxiv.org/abs/1904.08320} {arXiv:1904.08320 [nucl-th]} \BibitemShut
  {NoStop}%
\bibitem [{\citenamefont {Bellini}\ \emph {et~al.}(2021)\citenamefont
  {Bellini}, \citenamefont {Blum}, \citenamefont {Kalweit},\ and\ \citenamefont
  {Puccio}}]{Bellini:2020cbj}%
  \BibitemOpen
  \bibfield  {author} {\bibinfo {author} {\bibfnamefont {F.}~\bibnamefont
  {Bellini}}, \bibinfo {author} {\bibfnamefont {K.}~\bibnamefont {Blum}},
  \bibinfo {author} {\bibfnamefont {A.~P.}\ \bibnamefont {Kalweit}},\ and\
  \bibinfo {author} {\bibfnamefont {M.}~\bibnamefont {Puccio}},\ }\href
  {https://doi.org/10.1103/PhysRevC.103.014907} {\bibfield  {journal} {\bibinfo
   {journal} {Phys. Rev. C}\ }\textbf {\bibinfo {volume} {103}},\ \bibinfo
  {pages} {014907} (\bibinfo {year} {2021})},\ \Eprint
  {https://arxiv.org/abs/2007.01750} {arXiv:2007.01750 [nucl-th]} \BibitemShut
  {NoStop}%
\bibitem [{\citenamefont {Mahlein}\ \emph {et~al.}(2023)\citenamefont
  {Mahlein}, \citenamefont {Barioglio}, \citenamefont {Bellini}, \citenamefont
  {Fabbietti}, \citenamefont {Pinto}, \citenamefont {Singh},\ and\
  \citenamefont {Tripathy}}]{Mahlein:2023fmx}%
  \BibitemOpen
  \bibfield  {author} {\bibinfo {author} {\bibfnamefont {M.}~\bibnamefont
  {Mahlein}}, \bibinfo {author} {\bibfnamefont {L.}~\bibnamefont {Barioglio}},
  \bibinfo {author} {\bibfnamefont {F.}~\bibnamefont {Bellini}}, \bibinfo
  {author} {\bibfnamefont {L.}~\bibnamefont {Fabbietti}}, \bibinfo {author}
  {\bibfnamefont {C.}~\bibnamefont {Pinto}}, \bibinfo {author} {\bibfnamefont
  {B.}~\bibnamefont {Singh}},\ and\ \bibinfo {author} {\bibfnamefont
  {S.}~\bibnamefont {Tripathy}},\ }\href
  {https://doi.org/10.1140/epjc/s10052-023-11972-3} {\bibfield  {journal}
  {\bibinfo  {journal} {Eur. Phys. J. C}\ }\textbf {\bibinfo {volume} {83}},\
  \bibinfo {pages} {804} (\bibinfo {year} {2023})},\ \Eprint
  {https://arxiv.org/abs/2302.12696} {arXiv:2302.12696 [hep-ex]} \BibitemShut
  {NoStop}%
\bibitem [{\citenamefont {Cleymans}\ \emph {et~al.}(2011)\citenamefont
  {Cleymans}, \citenamefont {Kabana}, \citenamefont {Kraus}, \citenamefont
  {Oeschler}, \citenamefont {Redlich},\ and\ \citenamefont {Sharma}}]{SHM1}%
  \BibitemOpen
  \bibfield  {author} {\bibinfo {author} {\bibfnamefont {J.}~\bibnamefont
  {Cleymans}}, \bibinfo {author} {\bibfnamefont {S.}~\bibnamefont {Kabana}},
  \bibinfo {author} {\bibfnamefont {I.}~\bibnamefont {Kraus}}, \bibinfo
  {author} {\bibfnamefont {H.}~\bibnamefont {Oeschler}}, \bibinfo {author}
  {\bibfnamefont {K.}~\bibnamefont {Redlich}},\ and\ \bibinfo {author}
  {\bibfnamefont {N.}~\bibnamefont {Sharma}},\ }\href
  {https://doi.org/10.1103/PhysRevC.84.054916} {\bibfield  {journal} {\bibinfo
  {journal} {Phys. Rev.}\ }\textbf {\bibinfo {volume} {C84}},\ \bibinfo {pages}
  {054916} (\bibinfo {year} {2011})},\ \Eprint
  {https://arxiv.org/abs/1105.3719} {arXiv:1105.3719 [hep-ph]} \BibitemShut
  {NoStop}%
\bibitem [{\citenamefont {Andronic}\ \emph {et~al.}(2011)\citenamefont
  {Andronic}, \citenamefont {Braun-Munzinger}, \citenamefont {Stachel},\ and\
  \citenamefont {St$\ddot{\mathrm{o}}$cker}}]{SHM2}%
  \BibitemOpen
  \bibfield  {author} {\bibinfo {author} {\bibfnamefont {A.}~\bibnamefont
  {Andronic}}, \bibinfo {author} {\bibfnamefont {P.}~\bibnamefont
  {Braun-Munzinger}}, \bibinfo {author} {\bibfnamefont {J.}~\bibnamefont
  {Stachel}},\ and\ \bibinfo {author} {\bibfnamefont {H.}~\bibnamefont
  {St$\ddot{\mathrm{o}}$cker}},\ }\href
  {https://doi.org/10.1016/j.physletb.2011.01.053} {\bibfield  {journal}
  {\bibinfo  {journal} {Phys. Lett.}\ }\textbf {\bibinfo {volume} {B697}},\
  \bibinfo {pages} {203} (\bibinfo {year} {2011})},\ \Eprint
  {https://arxiv.org/abs/1010.2995} {arXiv:1010.2995 [nucl-th]} \BibitemShut
  {NoStop}%
\bibitem [{\citenamefont {Becattini}\ \emph {et~al.}(2014)\citenamefont
  {Becattini}, \citenamefont {Grossi}, \citenamefont {Bleicher}, \citenamefont
  {Steinheimer},\ and\ \citenamefont {Stock}}]{SHM3}%
  \BibitemOpen
  \bibfield  {author} {\bibinfo {author} {\bibfnamefont {F.}~\bibnamefont
  {Becattini}}, \bibinfo {author} {\bibfnamefont {E.}~\bibnamefont {Grossi}},
  \bibinfo {author} {\bibfnamefont {M.}~\bibnamefont {Bleicher}}, \bibinfo
  {author} {\bibfnamefont {J.}~\bibnamefont {Steinheimer}},\ and\ \bibinfo
  {author} {\bibfnamefont {R.}~\bibnamefont {Stock}},\ }\href
  {https://doi.org/10.1103/PhysRevC.90.054907} {\bibfield  {journal} {\bibinfo
  {journal} {Phys. Rev.}\ }\textbf {\bibinfo {volume} {C90}},\ \bibinfo {pages}
  {054907} (\bibinfo {year} {2014})},\ \Eprint
  {https://arxiv.org/abs/1405.0710} {arXiv:1405.0710 [nucl-th]} \BibitemShut
  {NoStop}%
\bibitem [{\citenamefont {Vovchenko}\ and\ \citenamefont
  {St$\ddot{\mathrm{o}}$cker}(2017)}]{SHM4}%
  \BibitemOpen
  \bibfield  {author} {\bibinfo {author} {\bibfnamefont {V.}~\bibnamefont
  {Vovchenko}}\ and\ \bibinfo {author} {\bibfnamefont {H.}~\bibnamefont
  {St$\ddot{\mathrm{o}}$cker}},\ }\href
  {https://doi.org/10.1103/PhysRevC.95.044904} {\bibfield  {journal} {\bibinfo
  {journal} {Phys. Rev.}\ }\textbf {\bibinfo {volume} {C95}},\ \bibinfo {pages}
  {044904} (\bibinfo {year} {2017})},\ \Eprint
  {https://arxiv.org/abs/1606.06218} {arXiv:1606.06218 [hep-ph]} \BibitemShut
  {NoStop}%
\bibitem [{\citenamefont {Andronic}\ \emph {et~al.}(2018)\citenamefont
  {Andronic}, \citenamefont {Braun-Munzinger}, \citenamefont {Redlich},\ and\
  \citenamefont {Stachel}}]{SHM5}%
  \BibitemOpen
  \bibfield  {author} {\bibinfo {author} {\bibfnamefont {A.}~\bibnamefont
  {Andronic}}, \bibinfo {author} {\bibfnamefont {P.}~\bibnamefont
  {Braun-Munzinger}}, \bibinfo {author} {\bibfnamefont {K.}~\bibnamefont
  {Redlich}},\ and\ \bibinfo {author} {\bibfnamefont {J.}~\bibnamefont
  {Stachel}},\ }\href {https://doi.org/10.1038/s41586-018-0491-6} {\bibfield
  {journal} {\bibinfo  {journal} {Nature}\ }\textbf {\bibinfo {volume} {561}},\
  \bibinfo {pages} {321} (\bibinfo {year} {2018})},\ \Eprint
  {https://arxiv.org/abs/1710.09425} {arXiv:1710.09425 [nucl-th]} \BibitemShut
  {NoStop}%
\bibitem [{\citenamefont {Sharma}\ \emph {et~al.}(2019)\citenamefont {Sharma},
  \citenamefont {Cleymans}, \citenamefont {Hippolyte},\ and\ \citenamefont
  {Paradza}}]{SHM6}%
  \BibitemOpen
  \bibfield  {author} {\bibinfo {author} {\bibfnamefont {N.}~\bibnamefont
  {Sharma}}, \bibinfo {author} {\bibfnamefont {J.}~\bibnamefont {Cleymans}},
  \bibinfo {author} {\bibfnamefont {B.}~\bibnamefont {Hippolyte}},\ and\
  \bibinfo {author} {\bibfnamefont {M.}~\bibnamefont {Paradza}},\ }\href
  {https://doi.org/10.1103/PhysRevC.99.044914} {\bibfield  {journal} {\bibinfo
  {journal} {Phys. Rev.}\ }\textbf {\bibinfo {volume} {C99}},\ \bibinfo {pages}
  {044914} (\bibinfo {year} {2019})},\ \Eprint
  {https://arxiv.org/abs/1811.00399} {arXiv:1811.00399 [hep-ph]} \BibitemShut
  {NoStop}%
\bibitem [{\citenamefont {Bass}\ \emph {et~al.}(2000)\citenamefont {Bass},
  \citenamefont {Danielewicz},\ and\ \citenamefont {Pratt}}]{Bass:2000az}%
  \BibitemOpen
  \bibfield  {author} {\bibinfo {author} {\bibfnamefont {S.~A.}\ \bibnamefont
  {Bass}}, \bibinfo {author} {\bibfnamefont {P.}~\bibnamefont {Danielewicz}},\
  and\ \bibinfo {author} {\bibfnamefont {S.}~\bibnamefont {Pratt}},\ }\href
  {https://doi.org/10.1103/PhysRevLett.85.2689} {\bibfield  {journal} {\bibinfo
   {journal} {Phys. Rev. Lett.}\ }\textbf {\bibinfo {volume} {85}},\ \bibinfo
  {pages} {2689} (\bibinfo {year} {2000})},\ \Eprint
  {https://arxiv.org/abs/nucl-th/0005044} {arXiv:nucl-th/0005044} \BibitemShut
  {NoStop}%
\bibitem [{\citenamefont {Acharya}\ \emph {et~al.}(2024)\citenamefont {Acharya}
  \emph {et~al.}}]{ALICE:2023asw}%
  \BibitemOpen
  \bibfield  {author} {\bibinfo {author} {\bibfnamefont {S.}~\bibnamefont
  {Acharya}} \emph {et~al.} (\bibinfo {collaboration} {ALICE}),\ }\href
  {https://doi.org/10.1007/JHEP09(2024)102} {\bibfield  {journal} {\bibinfo
  {journal} {JHEP}\ }\textbf {\bibinfo {volume} {09}},\ \bibinfo {pages}
  {102}},\ \Eprint {https://arxiv.org/abs/2308.16706} {arXiv:2308.16706
  [hep-ex]} \BibitemShut {NoStop}%
\bibitem [{\citenamefont {Pruneau}\ \emph {et~al.}(2024)\citenamefont
  {Pruneau}, \citenamefont {Basu}, \citenamefont {Gonzalez}, \citenamefont
  {Hanley}, \citenamefont {Marin}, \citenamefont {Dobrin},\ and\ \citenamefont
  {Manea}}]{Pruneau:2024jpa}%
  \BibitemOpen
  \bibfield  {author} {\bibinfo {author} {\bibfnamefont {C.}~\bibnamefont
  {Pruneau}}, \bibinfo {author} {\bibfnamefont {S.}~\bibnamefont {Basu}},
  \bibinfo {author} {\bibfnamefont {V.}~\bibnamefont {Gonzalez}}, \bibinfo
  {author} {\bibfnamefont {B.}~\bibnamefont {Hanley}}, \bibinfo {author}
  {\bibfnamefont {A.}~\bibnamefont {Marin}}, \bibinfo {author} {\bibfnamefont
  {A.~F.}\ \bibnamefont {Dobrin}},\ and\ \bibinfo {author} {\bibfnamefont
  {A.}~\bibnamefont {Manea}},\ }\href
  {https://doi.org/10.1103/PhysRevC.109.064913} {\bibfield  {journal} {\bibinfo
   {journal} {Phys. Rev. C}\ }\textbf {\bibinfo {volume} {109}},\ \bibinfo
  {pages} {064913} (\bibinfo {year} {2024})},\ \Eprint
  {https://arxiv.org/abs/2403.13007} {arXiv:2403.13007 [hep-ph]} \BibitemShut
  {NoStop}%
\bibitem [{\citenamefont {Bierlich}\ and\ \citenamefont
  {Christiansen}(2025)}]{Bierlich:2025pkg}%
  \BibitemOpen
  \bibfield  {author} {\bibinfo {author} {\bibfnamefont {C.}~\bibnamefont
  {Bierlich}}\ and\ \bibinfo {author} {\bibfnamefont {P.}~\bibnamefont
  {Christiansen}},\ }\href@noop {} {\  (\bibinfo {year} {2025})},\ \Eprint
  {https://arxiv.org/abs/2506.18375} {arXiv:2506.18375 [hep-ph]} \BibitemShut
  {NoStop}%
\bibitem [{\citenamefont {Andersson}\ \emph {et~al.}(1983)\citenamefont
  {Andersson}, \citenamefont {Gustafson}, \citenamefont {Ingelman},\ and\
  \citenamefont {Sjostrand}}]{Andersson:1983ia}%
  \BibitemOpen
  \bibfield  {author} {\bibinfo {author} {\bibfnamefont {B.}~\bibnamefont
  {Andersson}}, \bibinfo {author} {\bibfnamefont {G.}~\bibnamefont
  {Gustafson}}, \bibinfo {author} {\bibfnamefont {G.}~\bibnamefont
  {Ingelman}},\ and\ \bibinfo {author} {\bibfnamefont {T.}~\bibnamefont
  {Sjostrand}},\ }\href {https://doi.org/10.1016/0370-1573(83)90080-7}
  {\bibfield  {journal} {\bibinfo  {journal} {Phys. Rept.}\ }\textbf {\bibinfo
  {volume} {97}},\ \bibinfo {pages} {31} (\bibinfo {year} {1983})}\BibitemShut
  {NoStop}%
\bibitem [{\citenamefont {Dal}\ and\ \citenamefont
  {Raklev}(2015)}]{Dal:2015sha}%
  \BibitemOpen
  \bibfield  {author} {\bibinfo {author} {\bibfnamefont {L.~A.}\ \bibnamefont
  {Dal}}\ and\ \bibinfo {author} {\bibfnamefont {A.~R.}\ \bibnamefont
  {Raklev}},\ }\href {https://doi.org/10.1103/PhysRevD.91.123536} {\bibfield
  {journal} {\bibinfo  {journal} {Phys. Rev. D}\ }\textbf {\bibinfo {volume}
  {91}},\ \bibinfo {pages} {123536} (\bibinfo {year} {2015})},\ \bibinfo {note}
  {[Erratum: Phys.Rev.D 92, 069903 (2015), Erratum: Phys.Rev.D 92, 089901
  (2015)]},\ \Eprint {https://arxiv.org/abs/1504.07242} {arXiv:1504.07242
  [hep-ph]} \BibitemShut {NoStop}%
\bibitem [{\citenamefont {Vovchenko}\ and\ \citenamefont
  {Stoecker}(2019)}]{Vovchenko:2019pjl}%
  \BibitemOpen
  \bibfield  {author} {\bibinfo {author} {\bibfnamefont {V.}~\bibnamefont
  {Vovchenko}}\ and\ \bibinfo {author} {\bibfnamefont {H.}~\bibnamefont
  {Stoecker}},\ }\href {https://doi.org/10.1016/j.cpc.2019.06.024} {\bibfield
  {journal} {\bibinfo  {journal} {Comput. Phys. Commun.}\ }\textbf {\bibinfo
  {volume} {244}},\ \bibinfo {pages} {295} (\bibinfo {year} {2019})},\ \Eprint
  {https://arxiv.org/abs/1901.05249} {arXiv:1901.05249 [nucl-th]} \BibitemShut
  {NoStop}%
\bibitem [{\citenamefont {Acharya}\ \emph
  {et~al.}(2020{\natexlab{d}})\citenamefont {Acharya} \emph
  {et~al.}}]{ALICE:2020nkc}%
  \BibitemOpen
  \bibfield  {author} {\bibinfo {author} {\bibfnamefont {S.}~\bibnamefont
  {Acharya}} \emph {et~al.} (\bibinfo {collaboration} {ALICE}),\ }\href
  {https://doi.org/10.1140/epjc/s10052-020-8125-1} {\bibfield  {journal}
  {\bibinfo  {journal} {Eur. Phys. J. C}\ }\textbf {\bibinfo {volume} {80}},\
  \bibinfo {pages} {693} (\bibinfo {year} {2020}{\natexlab{d}})},\ \Eprint
  {https://arxiv.org/abs/2003.02394} {arXiv:2003.02394 [nucl-ex]} \BibitemShut
  {NoStop}%
\bibitem [{\citenamefont {Acharya}\ \emph {et~al.}(2025)\citenamefont {Acharya}
  \emph {et~al.}}]{ALICE:2024rnr}%
  \BibitemOpen
  \bibfield  {author} {\bibinfo {author} {\bibfnamefont {S.}~\bibnamefont
  {Acharya}} \emph {et~al.} (\bibinfo {collaboration} {ALICE}),\ }\href
  {https://doi.org/10.1103/PhysRevLett.134.022303} {\bibfield  {journal}
  {\bibinfo  {journal} {Phys. Rev. Lett.}\ }\textbf {\bibinfo {volume} {134}},\
  \bibinfo {pages} {022303} (\bibinfo {year} {2025})},\ \Eprint
  {https://arxiv.org/abs/2405.19890} {arXiv:2405.19890 [nucl-ex]} \BibitemShut
  {NoStop}%
\bibitem [{\citenamefont {Vovchenko}\ \emph {et~al.}(2019)\citenamefont
  {Vovchenko}, \citenamefont {D\"onigus},\ and\ \citenamefont
  {Stoecker}}]{Vovchenko:2019kes}%
  \BibitemOpen
  \bibfield  {author} {\bibinfo {author} {\bibfnamefont {V.}~\bibnamefont
  {Vovchenko}}, \bibinfo {author} {\bibfnamefont {B.}~\bibnamefont
  {D\"onigus}},\ and\ \bibinfo {author} {\bibfnamefont {H.}~\bibnamefont
  {Stoecker}},\ }\href {https://doi.org/10.1103/PhysRevC.100.054906} {\bibfield
   {journal} {\bibinfo  {journal} {Phys. Rev. C}\ }\textbf {\bibinfo {volume}
  {100}},\ \bibinfo {pages} {054906} (\bibinfo {year} {2019})},\ \Eprint
  {https://arxiv.org/abs/1906.03145} {arXiv:1906.03145 [hep-ph]} \BibitemShut
  {NoStop}%
\bibitem [{\citenamefont {Acharya}\ \emph
  {et~al.}(2021{\natexlab{a}})\citenamefont {Acharya} \emph
  {et~al.}}]{ALICE:2020jsh}%
  \BibitemOpen
  \bibfield  {author} {\bibinfo {author} {\bibfnamefont {S.}~\bibnamefont
  {Acharya}} \emph {et~al.} (\bibinfo {collaboration} {ALICE}),\ }\href
  {https://doi.org/10.1140/epjc/s10052-020-08690-5} {\bibfield  {journal}
  {\bibinfo  {journal} {Eur. Phys. J. C}\ }\textbf {\bibinfo {volume} {81}},\
  \bibinfo {pages} {256} (\bibinfo {year} {2021}{\natexlab{a}})},\ \Eprint
  {https://arxiv.org/abs/2005.11120} {arXiv:2005.11120 [nucl-ex]} \BibitemShut
  {NoStop}%
\bibitem [{\citenamefont {Acharya}\ \emph
  {et~al.}(2021{\natexlab{b}})\citenamefont {Acharya} \emph
  {et~al.}}]{ALICE:2020swj}%
  \BibitemOpen
  \bibfield  {author} {\bibinfo {author} {\bibfnamefont {S.}~\bibnamefont
  {Acharya}} \emph {et~al.} (\bibinfo {collaboration} {ALICE}),\ }\href
  {https://doi.org/10.1140/epjc/s10052-021-09349-5} {\bibfield  {journal}
  {\bibinfo  {journal} {Eur. Phys. J. C}\ }\textbf {\bibinfo {volume} {81}},\
  \bibinfo {pages} {630} (\bibinfo {year} {2021}{\natexlab{b}})},\ \Eprint
  {https://arxiv.org/abs/2009.09434} {arXiv:2009.09434 [nucl-ex]} \BibitemShut
  {NoStop}%
\bibitem [{\citenamefont {Acharya}\ \emph {et~al.}(2023)\citenamefont {Acharya}
  \emph {et~al.}}]{ALICE:2022xiu}%
  \BibitemOpen
  \bibfield  {author} {\bibinfo {author} {\bibfnamefont {S.}~\bibnamefont
  {Acharya}} \emph {et~al.} (\bibinfo {collaboration} {ALICE}),\ }\href
  {https://doi.org/10.1103/PhysRevLett.131.041901} {\bibfield  {journal}
  {\bibinfo  {journal} {Phys. Rev. Lett.}\ }\textbf {\bibinfo {volume} {131}},\
  \bibinfo {pages} {041901} (\bibinfo {year} {2023})},\ \Eprint
  {https://arxiv.org/abs/2204.10166} {arXiv:2204.10166 [nucl-ex]} \BibitemShut
  {NoStop}%
\bibitem [{\citenamefont {Vovchenko}(2024)}]{Vovchenko:2024pvk}%
  \BibitemOpen
  \bibfield  {author} {\bibinfo {author} {\bibfnamefont {V.}~\bibnamefont
  {Vovchenko}},\ }\href {https://doi.org/10.1103/PhysRevC.110.L061902}
  {\bibfield  {journal} {\bibinfo  {journal} {Phys. Rev. C}\ }\textbf {\bibinfo
  {volume} {110}},\ \bibinfo {pages} {L061902} (\bibinfo {year} {2024})},\
  \Eprint {https://arxiv.org/abs/2409.01397} {arXiv:2409.01397 [hep-ph]}
  \BibitemShut {NoStop}%
\end{thebibliography}%

\end{document}